\documentclass[a4paper,fleqn,usenatbib]{mnras}

\usepackage[T1]{fontenc}
\usepackage{ae,aecompl}

\usepackage{graphicx}	
\usepackage{amsmath}	
\usepackage{amssymb}	

\title[Long-term  monitoring of NGC7469]{Long-term optical spectral monitoring of NGC 7469}

\author[A. I. Shapovalova et al.]
{Alla I. Shapovalova,$^{1}$\thanks{E-mail: ashap@sao.ru}
L. \v C. Popovi\'c$^{2,3,4}$, V. H. Chavushyan$^{5}$, V. L. Afanasiev$^{1}$, \and
D. Ili\'c$^{3,4}$,  A. Kova\v cevi\'c$^{3,4}$, A. N. Burenkov$^{1}$, 
W. Kollatschny$^{6}$, O. Spiridonova$^{1}$, \and
J. R. Valdes$^{5}$,  N. G. Bochkarev$^{7}$,
V.  Patino-Alvarez$^{5}$, L. Carrasco$^{5}$ and V. E. Zhdanova$^{1}$
\\
$^{1}$Special Astrophysical Observatory of the Russian AS,
Nizhnij Arkhyz, Karachaevo-Cherkesia 369167, Russia\\
$^{2}$Astronomical Observatory, Volgina 7, 11160 Belgrade 74, Serbia\\
$^{3}$Department of Astronomy, Faculty of Mathematics, University
of Belgrade, Studentski trg 16, 11000 Belgrade, Serbia\\
$^{4}$Isaac Newton Institute of Chile, Yugoslavia Branch, Volgina 7, Belgrade, Serbia\\
$^{5}$Instituto Nacional de Astrof\'{\i}sica, \'{O}ptica y
Electr\'onica, Apartado Postal 51-216, 72000 Puebla, Puebla, M\'exico\\
$^{6}$Institut f\"ur Astrophysik, Georg-August-Universit\"at G\"ottingen, Germany\\
$^{7}$Sternberg Astronomical Institute, 119992 Moscow, Russia
}

\date{Accepted XXX. Received YYY; in original form ZZZ}

\pubyear{2016}

\begin{document}
\label{firstpage}
\pagerange{\pageref{firstpage}--\pageref{lastpage}}
\maketitle

\begin{abstract}
We present the results of the long-term (20-year period, from 1996 to 
2015) optical spectral monitoring of the Seyfert 1 galaxy NGC 7469. The variation in 
the light-curves of the broad He II $\lambda$4686\AA\, H$\beta$ and H$\alpha$ lines, and the continuum at 5100 \AA\ and 6300 \AA\  have been explored. The maximum of activity was in 1998, and the variability in the continuum and lines seems to have two periods of around 1200 and 2600 days, however these periodicities should be taken with caution because of the red-noise. Beside these periods, there are several short-term (1-5 days) flare-like events in the light-curves.
There are good correlations between the continuum fluxes and H$\alpha$ and H$\beta$ line fluxes, but significantly smaller correlation between the He II and continuum. We found that the time-lags between the continuum and broad lines are different for H$\beta$ ($\sim 20$ l.d.) and H$\alpha$ ($\sim 3$ l.d.), and that 
He II also has a smaller lag ($\sim$2-3 l.d.). The H$\alpha$ and H$\beta$ line profiles show a slight red asymmetry, and the broad line profiles did not changed in the 20-year period. 
Using the lags and widths of H$\alpha$ and H$\beta$ we estimated the central black hole mass and found that it is $\sim(1-6)\cdot 10^7\ M_\odot$, which is in agreement with previous reverberation estimates.

\end{abstract}




\section{Introduction}

The variability of Type 1 active galactic nuclei (AGNs) seems to be very common and can be used for investigation of the central part of an AGN, especially of the broad line region (BLR) which emits broad emission lines. A BLR of AGNs is expected to be gravitationally bounded by the central supermassive black hole, and consequently the gas in the BLR is expected to be virialized. This, knowing the broad line widths and the BLR dimension (from reverberation studies) can be used for the black hole mass determination
\citep[see e.g.][and references therein]{pet14b}. Therefore, the spectral monitoring of Type 1 AGNs with the variable broad emission line and continuum fluxes is very useful, not only for the black hole mass estimates, but also to explore the variability in the BLR and investigate the nature of the emission gas motion. One of the well-known Type 1 AGNs, that shows a significant variability in the continuum and line spectra, is
NGC 746. Its AGN activity has been monitored in different spectral bands
\citep[see e.g.][]{dul92,lei96,pro97,wan97,col98,nan98,pro09,per09,art10,dor10,ugo11,pet14,bal15}

NGC 7469 is probably interacting with another galaxy, IC 5283, that is about
60-70 Mpc far away \citep[see, e.g.][]{mar94}. Around the active nucleus, a bright circumnuclear 
ring is located at an angular distance of 2.5 arcsec (or $\sim$1 kpc) from the nucleus. The stellar ring shows 
star-forming activity \citep[starburst region, see][]{wil86,wil91,dia07,iz15}, with possible supernova explosions. E.g. the supernova SN 2000ft has been  observed in the circumnuclear region in the radio \citep{col01,per09} and optical light \citep{col07}

The spectrum of NGC 7469 shows broad emission lines superposed with strong narrow lines, which is typical for Seyfert 1 galaxies. However, one can in addition expect a contribution of the H II region emission from the star-forming ring. 

As was noted above, the AGN of NGC 7469 has been monitored in the X, UV and optical spectral bands. 
The emission lines (H$\beta$ and He II $\lambda$4686\AA ) as well as the continuum are variable. 
However, previous spectral monitoring campaigns covered mostly shorter time periods (of the order
of several months). In this paper we present photometric and spectroscopic observations of NGC 7469,
with the aim to analyse the long-term variability in the broad lines and continuum fluxes, and in the
broad line profiles. For the first time, we present the 20-year long monitoring campaign of  NGC 7469.

The paper is organized as following: In section 2 we describe the observations and reduction of observed data, in section 3 we analyse the observed photometric and spectral data and give results, in section 4 the obtained results are discussed, and finally in section 5 we outline our main conclusions.

\begin{figure}
\centering
\includegraphics[width=\columnwidth]{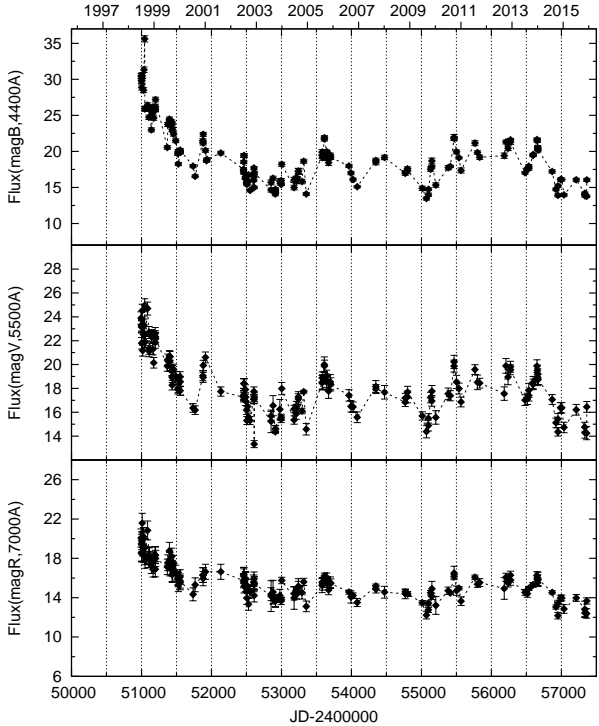}
\caption{Light-curves of the photometry magnitudes in the B, V, R filters, transformed into corresponding fluxes using the equations from \citet{di98}. The fluxes are given in units of 10$^{-15}$ erg cm$^{-2}$s$^{-1}$\AA$^{-1}$.} \label{fig1}
\end{figure}

\begin{table*}
\centering
\caption{The measured photometric magnitudes. Columns are: (1): Number, (2): UT date, (3): Modified Julian date (MJD), (4): Mean seeing in arcsec, and (5)-(7): BRV magnitudes and corresponding errors. The full table is available online.}
\label{tab1}
\begin{tabular}{clccccc}
\hline
N & UT-date& MJD & Seeing & M$_{\rm B}$ $\pm \sigma$ & M$_{\rm V}$ $\pm \sigma$ & M$_{\rm R}$ $\pm \sigma$  \\
  & YYMMDD & 2400000+ & [arcsec] &  &    &   \\
1 & 2 & 3 & 4 & 5 & 6 & 7 \\
\hline
  1  &  980629 & 50994.00 &  2.2  &  13.340  $\pm$0.02  &   12.957  $\pm$0.02   & 12.346  $\pm$0.02   \\           
  2  &  980706 & 51001.00 &  3.2  &  13.363  $\pm$0.02  &   12.991  $\pm$0.02   & 12.435  $\pm$0.02   \\           
  3  &  980707 & 51002.00 &  3.2  &  13.351  $\pm$0.02  &   13.057  $\pm$0.02   & 12.430  $\pm$0.02   \\           
  4  &  980708 & 51003.00 &  3.5  &  13.365  $\pm$0.02  &   12.983  $\pm$0.02   & 12.375  $\pm$0.02   \\           
  5  &  980709 & 51004.00 &  3.2  &  13.399  $\pm$0.02  &   12.988  $\pm$0.02   & 12.361  $\pm$0.02   \\           
  6  &  980716 & 51011.00 &  2.5  &  13.339  $\pm$0.02  &   12.929  $\pm$0.02   & 12.267  $\pm$0.02   \\           
  7  &  980723 & 51018.00 &  2.0  &     -               &   13.088  $\pm$0.02   & 12.399  $\pm$0.02   \\           
  8  &  980803 & 51029.00 &  2.0  &  13.413  $\pm$0.02  &   13.022  $\pm$0.02   & 12.339  $\pm$0.02   \\           
  9  &  980809 & 51035.00 &  3.0  &  13.310  $\pm$0.02  &   12.984  $\pm$0.02   & 12.440  $\pm$0.02   \\           
 10  &  980819 & 51045.00 &  3.5  &  13.171  $\pm$0.02  &   12.909  $\pm$0.02   & 12.453  $\pm$0.02   \\           
\hline
\end{tabular}
\end{table*}

\section{Observations and data reduction}

\subsection{Photometry}

The BVR photometry of NGC 7469 was performed at the Special Astrophysical Observatory of the Russian Academy of 
Science (SAO RAS) using the 1-m (1998--2014) and 60-cm (1998--2003) Zeiss telescopes with CCD (530$\times$580) or CCD1K (in 1998--2003) or CCD2K (in 2004--2014).
The telescopes are equipped with an offset guided automatic photometer. The characteristics of photometer are given in \citet{am00} and will not be repeated here. The photometric system is similar to those of Johnson in B and V, and of Cousins in R \citep{co76}. We used the software described in \citet{vl93} to reduce the observations. Photometric standard stars from \citet{pen71} - in 1998--2003, and from \citet{dor05} - in 2004--2015  have been used. These stars are close to NGC 7469, therefore the effects due to differential air mass are negligible for internal calibration purposes. In Table~\ref{tab1} we give the photometric  BVR-magnitude. For the light-curves (Fig \ref{fig1}), the magnitudes (m(B), m(V), m(R)) were transformed into fluxes F(B), F(V) and F(R) in units of 
10$^{-15}$ erg cm$^{-2}$s$^{-1}$\AA$^{-1}$, using the equations from \citet{di98}:
\begin{align}
    \log F(B,\lambda4400\AA)=-0.4m(B) - 8.180,\\
    \log F(V,\lambda5500\AA)=-0.4m(V) - 8.439,\\
    \log F(R,\lambda7000\AA)=-0.4m(R) - 8.759.
	\label{eq:photo}
\end{align}

\begin{table*}
\centering
\caption{Spectroscopic observations information. Columns are: (1): Observatory, (2): Code assigned to each combination of telescope + equipment used throughout this paper, (3): Telescope aperture and spectrograph. 
(4): Projected spectrograph entrance apertures (slit width$\times$slit length in
arcsec), and (5): Focus of the telescope.}
\label{tab2}
\begin{tabular}{lcccc}
\hline
Observatory & Code & Tel.aperture + equipment & Aperture [arcsec] & Focus \\
1 & 2 & 3 & 4 & 5\\
\hline
SAO (Russia)   & L(N)    & 6 m + Long slit  &  2.0$\times$6.0   & Nasmith    \\
SAO (Russia)   & L(U)    & 6 m + UAGS       &  2.0$\times$6.0   & Prime      \\
GHO (M\'exico) & GHO     & 2.1 m + B\&C     &  2.5$\times$6.0   & Cassegrain   \\
SAO (Russia)   & Z1      & 1 m + UAGS       &  4.0$\times$19.8  & Cassegrain   \\
SAO (Russia)   & Z1      & 1 m + UAGS       &  8.0$\times$19.8  & Cassegrain   \\
SAO (Russia)   & Z2K     & 1 m + UAGS       &  4.0$\times$9.45  & Cassegrain   \\
\hline
\end{tabular}
\end{table*}

\begin{table*}
\centering
\caption{Spectroscopic observations log. Columns are: (1): Number, (2): UT date, (3): Modified Julian date (JD), (4): CODE (Code given according to Table~\ref{tab2}.), (5): Projected
spectrograph entrance apertures, (6): Wavelength range covered, and (7): Mean seeing in arcsec. The full table is available online.}
\label{tab3}
\begin{tabular}{clccccc} 
\hline
N & UT-date & MJD & CODE & Aperture &Sp.range & Seeing  \\
  &    &2400000+ &  & [arcsec] & [\AA]   &  [arcsec] \\
1 & 2 & 3 & 4 & 5 & 6 & 7\\
\hline
  1 &  13.06.1996 &  50247.558 & L(U) &  2.0$\times$6.0    & 4400-5300  &  1.1   \\   
  2 &  14.06.1996 &  50248.550 & L(U) &  2.0$\times$6.0    & 4400-5300  &  1.4   \\
  3 &  12.07.1996 &  50276.535 & L(U) &  2.0$\times$6.0    & 3600-5400  &  1.1   \\
  4 &  13.07.1996 &  50277.528 & L(U) &  2.0$\times$6.0    & 4400-5300  &  1.6   \\
  5 &  16.07.1996 &  50280.567 & L(U) &  2.0$\times$6.0    & 4400-5300  &  1.2   \\
  6 &  17.07.1996 &  50281.576 & L(U) &  2.0$\times$6.0    & 3700-7150  &  2.8   \\
  7 &  25.07.1996 &  50289.575 & L(U) &  2.0$\times$6.0    & 4400-5300  &  3.6   \\
  8 &  10.08.1996 &  50305.567 & L(U) &  2.0$\times$6.0    & 4400-5300  &  3.2   \\
  9 &  08.09.1996 &  50335.447 & Z1   &  4.0$\times$19.8   & 4140-5800  &  3.0   \\
 10 &  11.09.1996 &  50338.422 & Z1   &  4.0$\times$13.5   & 3890-5450  &  5.0   \\
\hline
\end{tabular}
\end{table*}

\subsection{Spectral observations}
\label{sec:obs}

We monitored the galaxy over a period of 20 years, between June 13, 1996 
and July 23, 2015, collecting in total $\sim$260 nights of observation.
Spectra were acquired with the 6-m and 1-m telescopes of the SAO RAS, 
Russia (1996--2015), and with the INAOE's 2.1-m telescope of the ''Guillermo Haro Observatory'' (GHO) at Cananea, Sonora, M\'exico (1998--2007), using a long-slit spectrograph 
(UAGS in SAO RAS and Boller\&Chives in Mexico) equipped with a CCD. 
The typical wavelength interval covered was from $\sim$3800 \AA \ to $\sim$7400 \AA\, (see Table~\ref{tab3}), 
the spectral resolution was between 8-10 \AA \ or 12-15 \AA , and the signal-to-noise (S/N) ratio 
was S/N$>$50 in the continuum near the H$\beta$ line.

Observations with the 1-m SAO RAS telescope have been performed in two modes: 1) using CCD 2K$\times$2K
(EEV42-40,2068$\times$2072 pixels (pxs), 13.5$\times$13.5mkm), in the wavelength range from 3800 \AA\ to 
7400 \AA, with mean spectral resolution of $\sim$7.5 \AA\, and, where both H$\alpha$ and H$\beta$ lines are
covered with one spectrum. The normalization has been performed with the [O III] lines for both lines. 
The length scale along the slit was 1.35\arcsec\/px, totally taking 7pxs (9.45\arcsec), 
i.e. the aperture was 4.0\arcsec$\times$9.45\arcsec; and 2) in the period of 1996-2003 with  CCD 1k$\times$1k or 530$\times$580pxs; the slit width was 4.0\arcsec\ or 8.0\arcsec\  and the length scale along the slit was 2.2\arcsec/px, totally taking 9pxs (19.8\arcsec), i.e. the aperture was 4.0\arcsec(or 8.0\arcsec)$\times$19.8\arcsec. The spectral resolution was R$\sim$ 8-10 \AA . The blue (spectrum around H$\beta$) and red (spectrum around H$\alpha$) spectral bands have been observed separately.
However they are overlapping at the edges (red edge with blue and {\it vice versa}) 
for the wavelength interval of 200--400 \AA . 
This has been used for scaling the whole spectrum on [O III] lines. If some spectra were
poorly corrected for the spectral sensitivity at the ends, for the scaling of the red spectra the
doublet [SII]$\lambda\lambda$6717,6731 \AA\ has been used.

In the case of the 6-m SAO telescope the  spectrograph UAGS (prime focus) or a long slit (Nasmith  focus) with the aperture of 2.0\arcsec$\times$6.0\arcsec\ (slit length is 0.4\arcsec/px, taking totally 5 pxs). The  spectral resolution of this instrument is between 5 \AA\ and 8 \AA .

Observations on the 2.1-m (GHO) telescope were made with the aperture 2.5\arcsec$\times$6.0\arcsec\ 
(scale along slit: $\sim$ 0.46\arcsec/px; taking in total 13 pxs). Note here that from 1998 to 2004, 
the spectral observations  were done with a grism of 150 l/mm 
(a low dispersion of R=15 \AA ). From 2004 to 2007,  observations obtained
with a grism of 300 l/mm (a moderate dispersion of R=8.0 \AA) were added.
As a rule, the observations were performed with the moderate dispersion
in the blue (around H$\beta$) or red (around H$\alpha$) spectral band during the first night of each set and 
usually during the next night we used the low dispersion spectrograph
covering the whole spectral range from 4000 \AA\ to 7500 \AA.

Since the shape of the continuum of active galaxies practically
does not change during adjacent nights, it was easy to link together
the blue and red spectral bands obtained with the moderate spectral dispersion, using
the data obtained for the continuum with the low-dispersion
in the whole wavelength range. 

Spectrophotometric standard  stars were observed every night. 
Information on the source of spectroscopic observations is listed in Table~\ref{tab2}.
The log of spectroscopic observations is given in Table~\ref{tab3}.

The spectrophotometric data reduction was carried out using either the software
developed at SAO RAS \citep{vl93} or the IRAF package for the spectra obtained in Mexico.
The image reduction process included bias and flat-field corrections, cosmic 
ray removal, 2D wavelength linearization, sky spectrum subtraction, addition 
of the spectra for every night, and relative flux calibration based on standard
star observations. In the analysis, about  10\%  of the spectra were discarded 
for several different reasons (e.g. high noise level, badly corrected spectral 
sensitivity, poor spectral resolution $>$15 \AA, etc.). Thus our final data 
set consisted of 233 blue (covering He II and H$
\beta$) and 108 red (covering H$\alpha$) spectra, taken during 240 nights which 
we use in further analysis.

\begin{table}
\centering
\caption{Spectral flux scale factor $\varphi$ and extended source correction 
G(g) [in units of 10$^{-15} \rm erg \ cm^{-2} s^{-1}$\AA$^{-1}$] for 
different telescopes.}
\label{tab4}
\begin{tabular}{lccc}
\hline
Sample &Aperture& Scale factor& Extended source  correction \\
  &  (arcsec) & ($\varphi$) & G(g) \\
\hline
  Z2K      &     4.0$\times$9.45  &  1.000            &   0.000      \\
  Z1       &     8.0$\times$19.8  &  1.069$\pm$0.024  &   3.62$\pm$0.24 \\
  Z1       &     4.0$\times$19.8  &  1.017$\pm$0.027  &   1.58$\pm$0.74 \\
  L(U,N)   &     2.0$\times$6.0   &  1.000            &  -2.0        \\ 
  GHO      &     2.5$\times$6.0   &  0.967$\pm$0.012  &  -2.52$\pm$0.82 \\
\hline
\end{tabular}
\end{table}

\subsection{Absolute calibration (scaling) of the spectra}
\label{sec:cal}

The standard technique of the flux calibration 
based on comparison with the stars of known spectral energy distribution
is not precisely enough for studying the AGN spectral variability. Consequently
 for the absolute calibration, we used fluxes of the narrow emission 
lines  to scale the AGN spectra, since one cannot expect that 
they vary in a period of tens of years \citep{pet93}.
However, recently \cite{pet13} 
found  narrow-line variability in NGC 5548 on a shorter timescale. 
In the case of NGC 7469  the flux of  the [O III]$\lambda$\,5007\AA\ line has not changed noticeably 
between 1996 \citep{col98} 
and 2010 \citep{pet14},
i.e. [O III]$\lambda$5007\AA\ flux is in general agreement between two epochs with time distance of 14 years. Therefore we assume that the flux of the [O III]$\lambda$\,5007\AA\ line in NGC 7469 remained 
constant during  our monitoring period (20 years).

All blue spectra of NGC 7469 were thus scaled to the constant flux \- $F$([O III]$\lambda\,5007\AA )= 6.14\times 10^{-13}$\,erg\,s$^{-1}$\,cm$^{-2}$  determined by \cite{pet14}.
The scaling method  of the blue spectra have been described in several papers 
\citep[see][and references therein]{sh04,sh10,sh12,sh16}
and will not be repeated here. This method allows to obtain a homogeneous set of spectra with the same 
wavelength calibration and the same [O III]$\lambda$5007\AA\ flux.

The spectra obtained using the SAO 1-m telescope with the
mean resolution of 7.5\,\AA\ (UAGS+CCD2K, Table~\ref{tab2}) are covering  both 
the H$\alpha$ and H$\beta$ spectral bands. These spectra were scaled using the
[O III]$\lambda$\,5007\AA\ line, and the red spectral band
was automatically scaled  (also by the [O III]$\lambda$\,5007\AA\ flux).

Blue spectra, taken with the 1-m and 6-m telescopes with CCD 1K$\times$1K or 530$\times$580pxs, and with the 2.1-m telescope with a grism of 300 l/mm  (Code L(N),L(U),Z1, and GHO from Tables 2 and 3)  were also scaled  using the [O III]$\lambda$5007\AA\ line. In the same night, the scaled blue spectra was merged with  the corresponding red spectrum using the overlapping portions of this  spectra (see section 2.2).  However, the accuracy of the scaling procedure depends strongly on the determination of the continuum slope in the red spectral band, i.e. one has to carefully account the spectral sensitivity of the equipment. This has been performed by using the stars for comparison. In poor photometric conditions (clouds, mist, etc.) the reduction can give a wrong spectral slope (fall or rise) and, consequently, the errors in the scaling procedure for the H$\alpha$ wavelength band can be larger. In the case that only the red spectral band was observed, we used the flux of doublets [S II]$\lambda\lambda$6717,6731\AA\ for the absolute calibration. Another source of uncertainty is the fact that the red wing of the [S II]$\lambda$6731\AA\ narrow line overlaps with the atmospheric B band (6870\AA), which cannot be properly removed from the end of the red spectral band. Taking all the above facts, we exclude 12 red spectra from the analysis. Most of the H$\alpha$ fluxes have been determined from the spectra that were scaled using the [O III] narrow line. Only 7 red spectra have been scaled using the [S II]$\lambda\lambda$6717,6731\AA\ narrow lines.

\begin{figure*}
\centering
\includegraphics[width=10cm,angle=90]{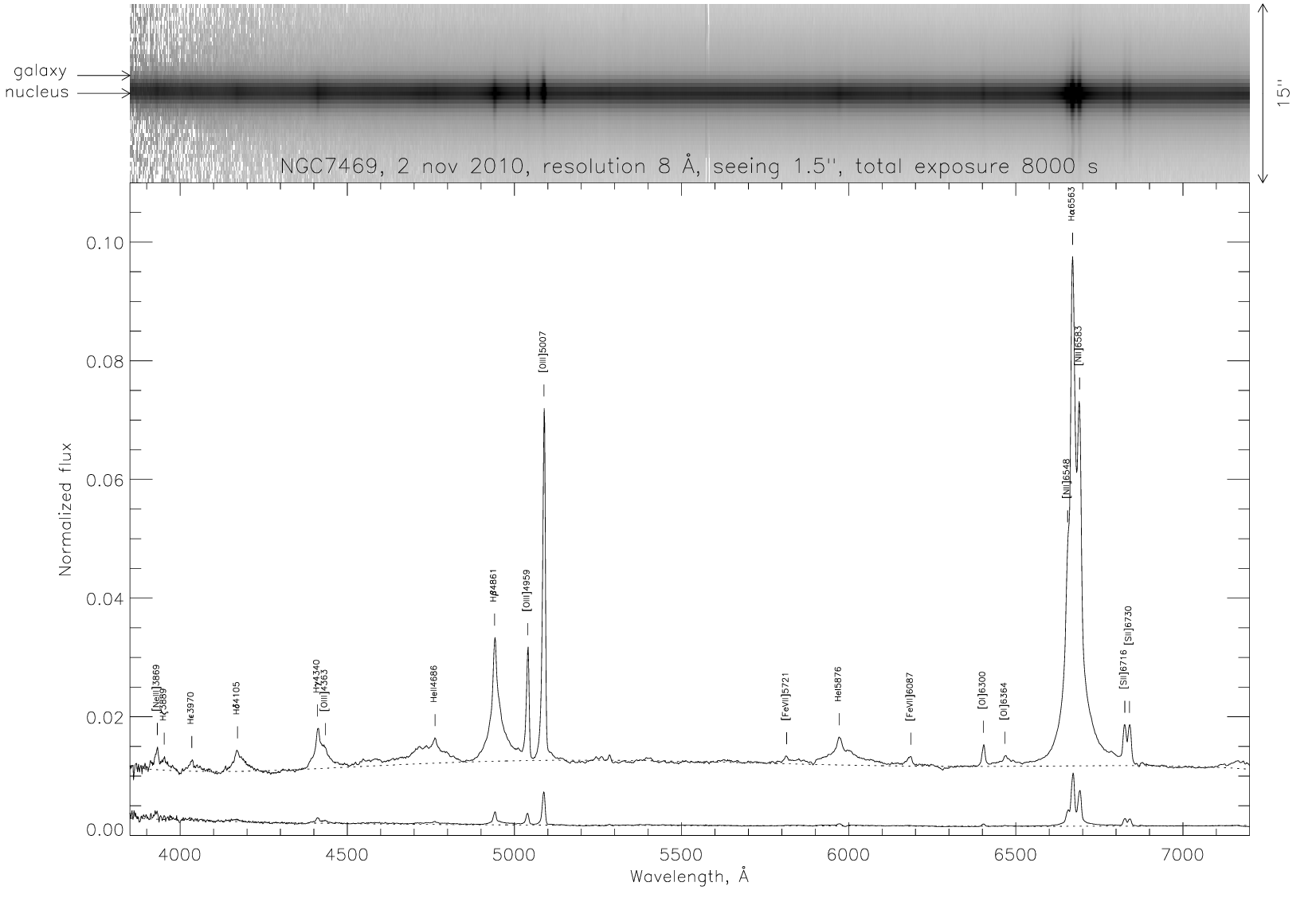}
\includegraphics[width=10cm,angle=90]{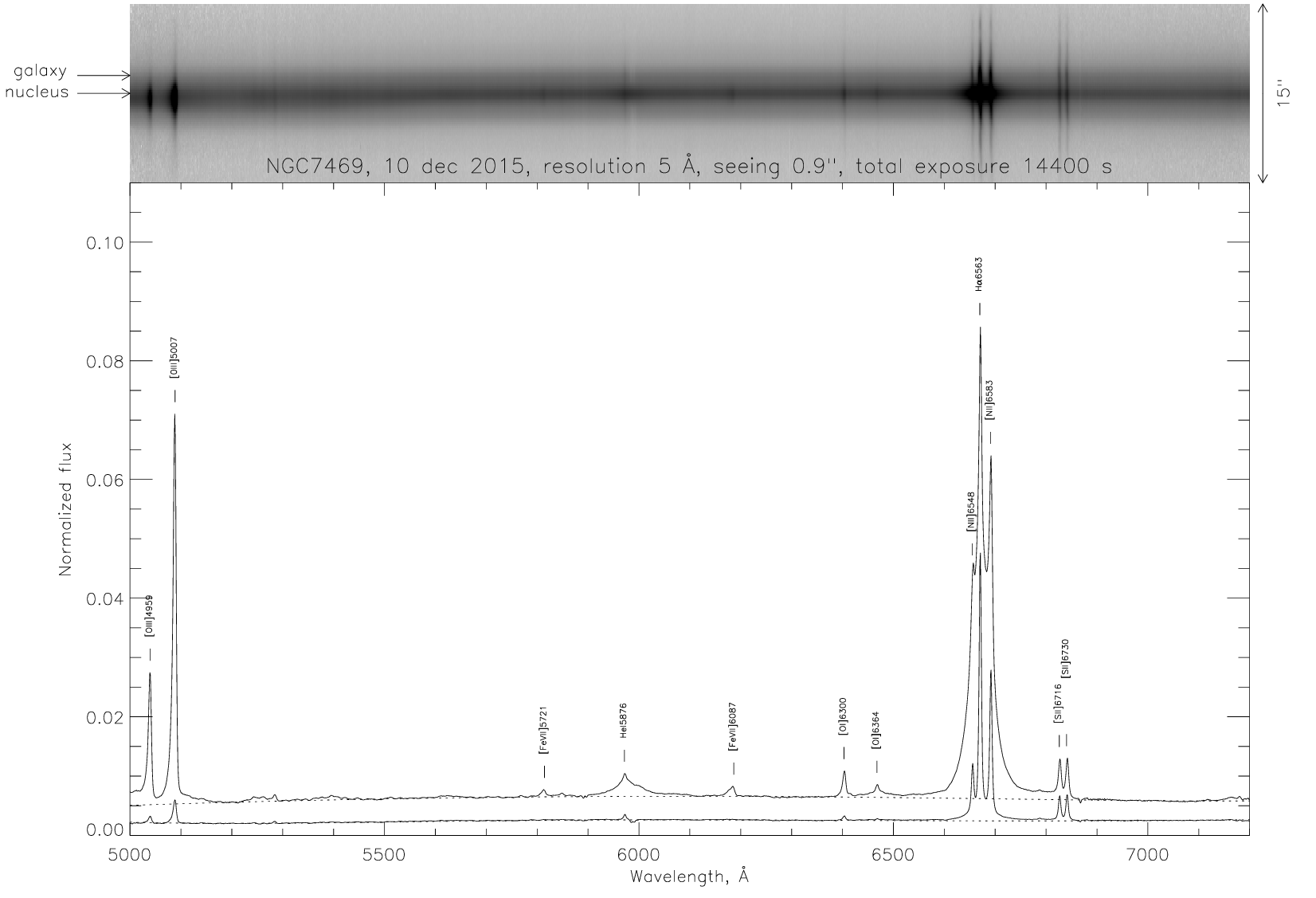}
\caption{Comparison of the AGN to the surrounding stellar-ring emission, using the spectra from two epochs with spectral 
resolution of 8 \AA\ (upper panel) and 5 \AA\ (bottom panel). The arrows on the spectrum images 
(top sub-panels) show the positions of the AGN and stellar-ring emission, for which the corresponding spectra are shown below. The strongest emission lines are denoted.} \label{fig-s}
\end{figure*}

\subsection{Unification of the spectral data}
\label{sec:uni}

In order to investigate the long-term spectral variability of an AGN, 
it is necessary to gather a consistent and uniformed data set.  
Since observations were carried out using instruments of different apertures, 
it is necessary to correct the line and continuum fluxes for these 
effects \citep{pc83}. As in our previous 
papers \citep{sh01, sh04,sh08, sh10, sh12, sh13,sh16},
we determined a point-source correction factor 
($\varphi$) and an aperture-dependent correction factor to 
account for the host galaxy contribution to the continuum (G(g)), and 
used the following expressions \citep[see][]{pet95}:
\begin{align}
    F({\rm line}){\rm true} = \varphi \cdot F({\rm line}){\rm obs},\\   
    F({\rm cont}){\rm true} = \varphi \cdot  F({\rm cont}){\rm obs} - G(g),
	\label{eq:aperture}
\end{align}
where index "obs" denotes the observed flux, and "true" the aperture corrected flux.
The spectra of the 1-m telescope+UAGS+CCD2K, within an aperture of
4\arcsec $\times$ 9.45\arcsec\ were adopted as standard (i.e. $\varphi$= 1.0, 
G(g)=0 by definition). 
The correction factors $\varphi$ and G(g) are determined empirically 
by comparing pairs of simultaneous observations from each of given telescope 
data sets to that of the standard data set \citep[as it was used in AGN Watch, see e.g.][]
{pet94,pet98,pet02}. The time intervals between observations which have been defined 
as "nearly simultaneous" are typically of 1--3 days. Note here, that in cases of a higher difference ($>10$\%) between two simultaneous observations (in the 1--3 days interval), which may indicate a fast changing in the flux or some artificial effect, we considered them separately (see Sections~\ref{sec:star} and \ref{sec:flare}).
The point-source scale correction factor $\varphi$ and extended-source 
correction factor G(g) values (in units of
10$^{-15} \rm erg \, cm^{-2} s^{-1} \AA^{-1}$) for different samples are 
listed in Table~\ref{tab4}.

\subsection{Flux of the star-forming region vs. the AGN}
\label{sec:star}

For the determination of the observed fluxes of H$\alpha$ and H$\beta$, it is necessary to subtract the
underlying continuum, which contains,  in addition to the AGN continuum, the contribution from the host galaxy and from the star-forming ring around the AGN \citep[see][]{da04,dia07,rg08}. The contribution of the circumnuclear ring can 
contribute, not only in the continuum emission but also in the narrow lines. Since we used the narrow
[O III] lines for absolute calibration (see Section~\ref{sec:cal}), the position of the slit, as well as, a small
off-centre of the slit can affect our calibration. To check this, we used observations taken with the 
Spectral Camera with Optical Reducer for Photometric and Interferometric Observations (SCORPIO) spectropolarimeter (the detailed description of the observations will be given in Afanasiev et al. 2016, in preparation).

We have four observations with different position angle of the slit. We show the observations from two epochs with the spectral resolution of 8 \AA\ and 5 \AA\ in Fig.~\ref{fig-s}.
As it can be seen in Fig.~\ref{fig-s}, there is a contribution of the circumnuclear ring emission in the narrow
lines and in the continuum. We accepted that the contribution of the circumnuclear ring and host galaxy in the continuum is $Fgal.=(8.7\pm 0.9)\cdot10^{-15}\rm erg\ cm^{-2}s^{-1}A^{-1}$ given by \cite{be13}.
The contribution of the circumnuclear ring to the narrow lines is not significant, especially in  lower resolution spectra (see Fig.~\ref{fig-s}, upper panel).

\begin{figure}
\centering
\includegraphics[width=\columnwidth]{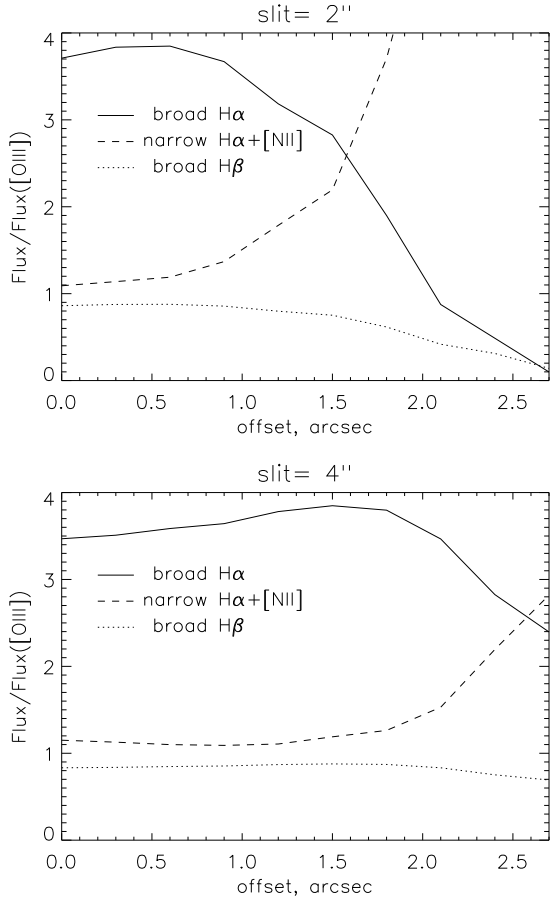}
\caption{Simulation of the slit offset influence on absolute line flux calibration with the [O III]
lines, using the slit of 2\arcsec\ (upper panel) and  4\arcsec\ (bottom panel).  Different lines are deonted on the plots.} \label{slit}
\end{figure}

\begin{figure}
\centering
\includegraphics[width=\columnwidth]{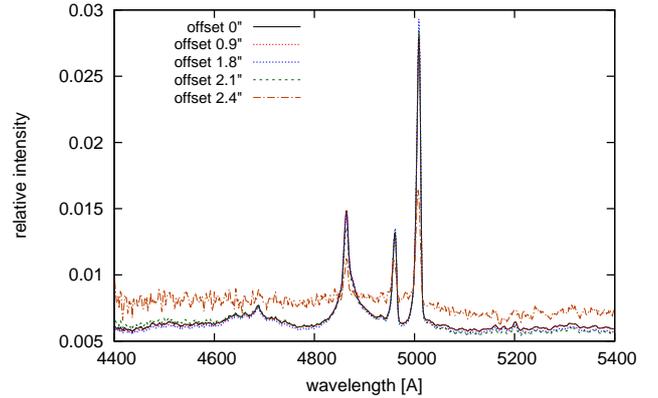}
\caption{Simulations of the observed spectra in the H$\beta$ spectral region (normalized to the [O III] lines) with different slit offsets (denoted in the upper left corner).}
\label{fig-off}
\end{figure}

However, a slight slit motion across the AGN and surrounding stellar ring may be present, and it can 
affect the intensity of the [O III] lines, and consequently give an artificial broad line variability. 
To test this, we explore the influence of the slit position to the calibration of the spectra using the [O III] lines. We simulated the influence of the slit of 2\arcsec\ (see Fig.~\ref{slit}, upper panel) moving across the field (presented in Fig.~\ref{fig-s}, upper spectrum image on panels), first taking that the AGN is in the centre (see arrows showing the AGN emission) and moving to the offset until 2.5\arcsec.
We measured the broad H$\alpha$ and H$\beta$, and narrow H$\alpha$+[N II] fluxes in the units of the [O III] lines. With such narrow slit, for offset $>$1.0\arcsec\,
we have a big influence of the slit position to the calibration of 
the line fluxes. Around 40\% of our spectra were obtained with the slit width of 2\arcsec\ (6-m telescope) and 2.5\arcsec\ 
(2.1-m telescope, see Table~\ref{tab2}), but for 60\% of our spectra the slit width of 
4.0\arcsec\ or 8.0\arcsec\ (1-m telescope) was used. Therefore we simulated a slit of 4\arcsec\ across the field of view, 
requesting that the AGN emission is in the slit. The result of this simulation is shown in Fig.~\ref{slit} (bottom panel), 
where one can see 
that more-or-less the broad H$\beta$ line has a constant intensity until the high offset of $<$2.0\arcsec, 
while the H$\alpha$ line tends to be more sensitive to the slit offset $>$1.0\arcsec. In Fig.~\ref{fig-off}
we present the H$\beta$ spectral region (normalized to the flux of [O III] lines) with different
offsets in the case of a slit of 4.0\arcsec. As can be seen in Fig.~\ref{fig-off}, 
until offset of $\sim 2$\arcsec\ the H$\beta$ spectral line, normalized to the [O III] lines flux, shows no major
change in the line flux and profile (it seems to be within the measured error-bars). But with offset larger 
than 2\arcsec\ (see the case of 2.4\arcsec\ offset in Fig.~\ref{fig-off}), there is a significant change in the broad line flux,
and in the continuum. 

From exploration of the above presented slit influence, one can conclude 
that the slit width of 2.0\arcsec\  moved  only for $>$1\arcsec\ from the central
position (i.e. the offset $>1$\arcsec, see Fig \ref{slit}, upper panel) can produce artificial, "flare-like" changes. If the width of the slit is 4\arcsec, one can see a similar phenomenon ("flare-like") in the H$\alpha$ flux for offset of $>$1.0\arcsec\ and in the H$\beta$ flux for offset of $>$2\arcsec\ (Fig.~\ref{slit}, bottom).

Taking into account the results of the slit motion simulation, we have reanalysed each "flare-like" event by comparing 
spectroscopic and photometric fluxes (aperture of photometric observations was 15\arcsec, i.e. we cannot expect the 
influence of the slit offset on these observations) and discarded those cases where a "flare-like" event is detected in the
spectroscopic light-curve, but absent in the photometric one.  As a result, we have removed from our analysis 12 spectra 
with the artificial "flare-like" events.

Taking into account the results of simulations we can conclude that the slit offset 
has more influence on H$\alpha$ than on H$\beta$, therefore in the further  analysis, we will concentrate more on the variability of the H$\beta$ line.

\subsection{Measurements of the spectral fluxes and errors}
\label{sec:flux}

From the scaled spectra (see Sections~\ref{sec:obs} and \ref{sec:cal})  we determined the averaged
flux in the  continuum  near the H$\beta$ line at the observed 
wavelength $\sim 5190$ \AA\, ($\sim 5100$ \AA\, in the rest frame),
by averaging fluxes in the 
spectral range of 5180--5200 \AA. The continuum near the H$\alpha$ 
line at the observed wavelength $\sim 6340$ \AA\, 
($\sim 6240$ \AA\, in the rest frame), has been measured by averaging  fluxes in the 
spectral range of 6330--6350 \AA. 

To measure the observed fluxes of H$\beta$ and H$\alpha$, 
it is necessary to subtract the underlying continuum. For this goal, 
a linear continuum has been fitted through the windows of 20\,\AA\, 
located at 4870\,\AA\, and 5140\,\AA\, for the H$\beta$, and 
at 6340\,\AA\, and  6970\,\AA\, 
for the H$\alpha$ spectral band. After the continuum subtraction, we defined 
the observed line fluxes in the following wavelength bands: 
from 4880\AA\  to 5012\,\AA\, for H$\beta$  and  from 6550 \AA\ to 6790 \,\AA\, for H$\alpha$. 
A linear continuum for He II has been taken  at 4505 \AA\ and 4860 \AA.
He II fluxes have been measured as fluxes between 4660 \AA\ and 4840 \AA . 

Using $\varphi$ and G(g) factors from Table~\ref{tab4}, we re-calibrated 
the observed H$\beta$ and H$\alpha$ fluxes, and their corresponding near-by continuum 
fluxes to a common scale using the standard aperture of 4.0\arcsec$\times$9.45\arcsec.
In Table~\ref{tab5}, the fluxes for the continua at the rest-frame wavelengths at 5100 \AA\, and 6240 \AA, as well as the He II, H$\beta$ and  H$\alpha$ lines and their errors are given. 
The mean errors (uncertainties) of the 
continuum fluxes at 5100 \AA\, and 6240 \AA\, as well as of the line fluxes of H$\beta$, and H$\alpha$  
are in the interval between  $\sim$ 2.3\% and 3.8\%, while for the He II line 
is estimated to be 6.4$\pm$4.4\% (see Table~\ref{tab6}). 
The error-bars have been estimated by comparing results from spectra 
obtained within the time interval that is shorter than 3 days.
The flux errors given  in Table~\ref{tab5}  were estimated by using the mean 
error given in Table~\ref{tab6}.

\begin{table*}
\centering
\caption{The measured continuum and line fluxes, and their estimated errors.
Columns are: (1): Number of spectra, (2): Observed date, (3): Julian Date in [2400000+], 
(4): Blue continuum, (5): H$\beta$, (6): Red continuum, (7): H$\alpha$, and
(8): He II. The line fluxes are in units of $10^{-13} \rm erg \, cm^{-2} s^{-1}$\AA$^{-1}$, and continuum fluxes
 in units of $10^{-15} \rm erg \, cm^{-2} s^{-1}$\AA$^{-1}$. The full table is available online.}
\label{tab5}
\begin{tabular}{clcccccc}
\hline
N & UT-date & JD+ & F${\rm 5100}\pm \sigma$ & F(H$\beta$)$\pm \sigma$ & F${\rm 6300}\pm \sigma$ & F(H$\alpha$)$\pm \sigma$  & F(He II)$\pm \sigma$ \\
1& 2 & 3 & 4&5& 6 & 7 & 8 \\
\hline
  1 & 13.06.1996 & 50247.56 & 14.94$\pm$0.48 &  8.14$\pm$0.31 &                &  & 2.71$\pm$0.17   \\
  2 & 14.06.1996 & 50248.55 & 15.50$\pm$0.50 &  8.49$\pm$0.32 &                &  & 2.33$\pm$0.15 \\
  3 & 12.07.1996 & 50276.54 & 13.54$\pm$0.44 &  8.00$\pm$0.30 &                &  & 2.04$\pm$0.42 \\
  4 & 13.07.1996 & 50277.53 & 14.20$\pm$0.46 &  8.20$\pm$0.31 &                &  & 2.58$\pm$0.53 \\
  5 & 16.07.1996 & 50280.57 & 14.59$\pm$0.47 &  7.95$\pm$0.30 &                &  & 1.72$\pm$0.35 \\
  6 & 17.07.1996 & 50281.58 & 16.41$\pm$0.53 &  8.01$\pm$0.30 &                &  &  \\
  7 & 25.07.1996 & 50289.58 & 16.94$\pm$0.55 &  7.50$\pm$0.28 &                &  &  \\
  8 & 10.08.1996 & 50305.57 & 18.73$\pm$0.61 &  8.14$\pm$0.31 &                &  & 1.52$\pm$0.10 \\
  9 & 08.09.1996 & 50335.45 & 19.60$\pm$0.63 &  8.94$\pm$0.34 &                &  &  \\
 10 & 11.09.1996 & 50338.42 & 18.50$\pm$0.60 &  9.97$\pm$0.37 &                &  &  \\
 11 & 12.09.1996 & 50339.49 & 19.07$\pm$0.62 &  9.76$\pm$0.37 &                &  &  \\
 12 & 02.11.1996 & 50390.17 & 18.03$\pm$0.58 & 11.02$\pm$0.41 &                &  & 3.67$\pm$0.24 \\
 13 & 25.08.1997 & 50685.55 & 22.32$\pm$0.72 & 11.69$\pm$0.44 &                &  &  \\
 14 & 27.08.1997 & 50688.45 & 24.05$\pm$0.78 & 11.32$\pm$0.42 &                &  & 4.33$\pm$0.28 \\
 15 & 28.08.1997 & 50689.50 & 21.83$\pm$0.71 & 12.05$\pm$0.45 &                &  & 4.04$\pm$0.26 \\
 16 & 30.08.1997 & 50691.49 & 22.87$\pm$0.74 & 12.15$\pm$0.46 &                &  & 4.77$\pm$0.31 \\
 17 & 09.09.1997 & 50701.45 & 23.39$\pm$0.76 & 11.85$\pm$0.44 & 19.03$\pm$0.57 & 48.18$\pm$1.12 & 3.82$\pm$0.24 \\
 18 & 10.09.1997 & 50702.39 & 23.68$\pm$0.77 & 12.19$\pm$0.46 & 18.87$\pm$0.57 & 47.68$\pm$1.11 & 4.11$\pm$0.26 \\
\hline
\end{tabular}
\end{table*}

\begin{table}
\centering
\caption{Estimates of the mean errors for continuum fluxes at 6200 and 5100\AA,
H$\alpha$, H$\beta$, and He II total-line fluxes. Columns are:
(1): Measured continuum or line, (2): Observed wavelength range,
(3): Rest-frame wavelength range for z=0.0164, and (4): Estimated error and its standard
deviation.}
\label{tab6}
\begin{tabular}{lcccc}
\hline
Cnt/Line &  Spectral Region &  Spectral Region    & $\sigma \pm$e \\
    &  [\AA] (obs)& [\AA] (rest) & [\%] \\
\hline
 cont 6200         &  6330-6350  & 6228-6248  & 3.02 $\pm$2.41  \\ 
 cont 5100         &  5180-5200  & 5096-5116  & 3.24 $\pm$2.30  \\
 H$\alpha$ - total &  6550-6790  & 6444-6680  & 2.32 $\pm$1.64   \\
 H$\beta$ - total  &  4880-5012  & 4801-4931  & 3.75 $\pm$2.75   \\
 He II - total     &  4660-4840  & 4585-4762  & 6.41 $\pm$4.38   \\
\hline
\end{tabular}
\end{table}

\begin{figure}
\centering
\includegraphics[width=\columnwidth]{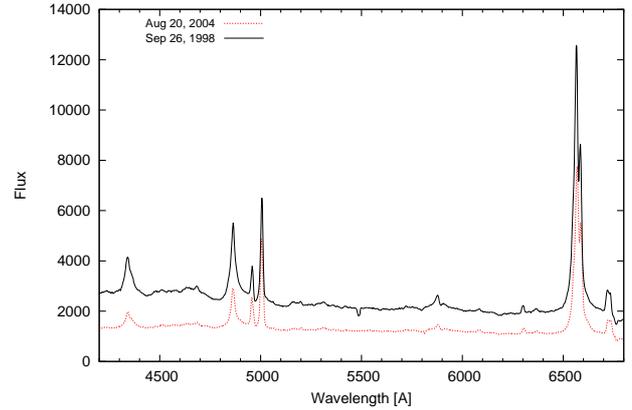}
\caption{Observed spectra in the minimum and maximum of activity during the monitoring period (epoch of observations denoted in the upper left corner).}
\label{fig-mm}
\end{figure}

\begin{figure*}
\centering
\includegraphics[width=12cm]{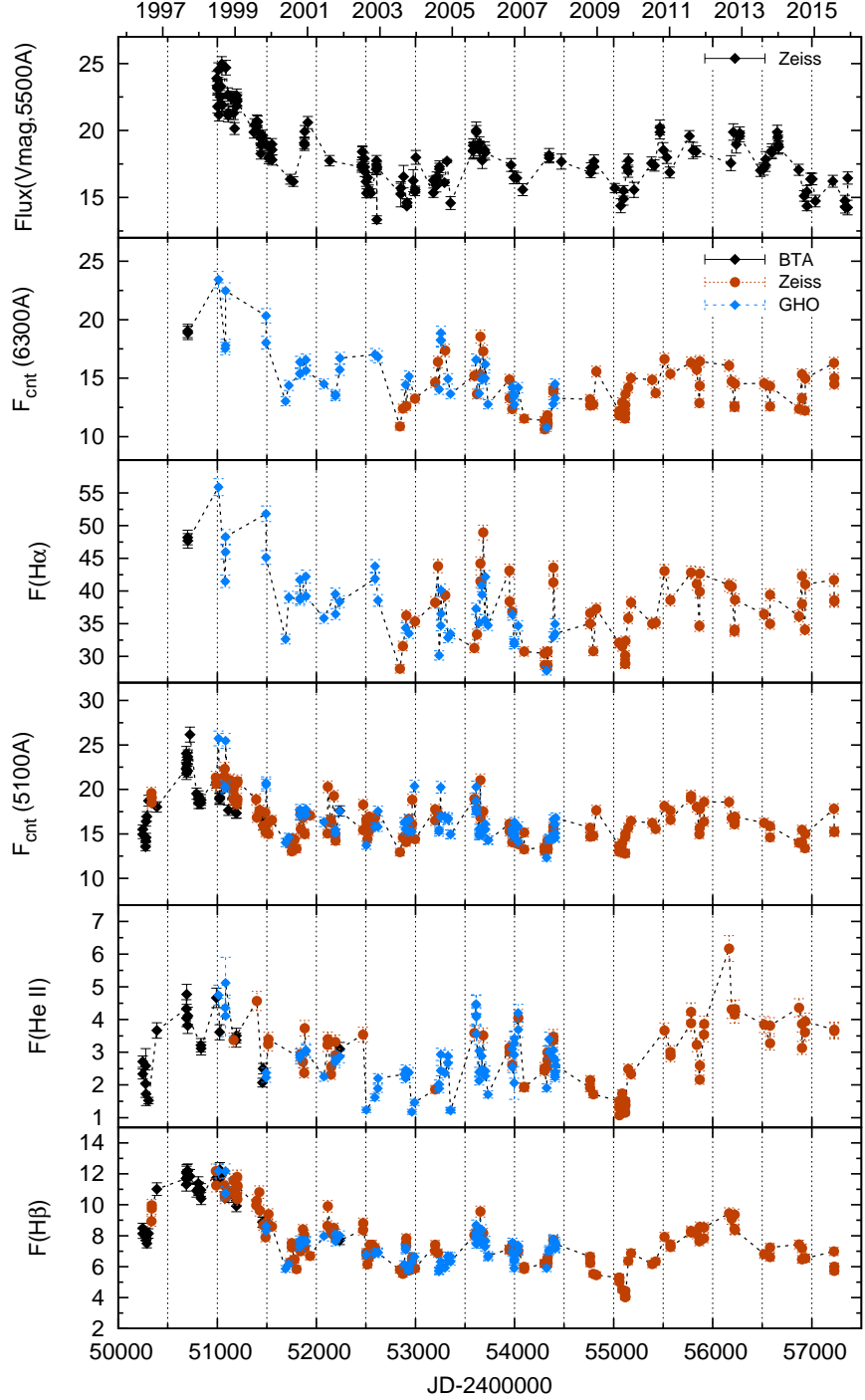}
\caption{Light-curves for the spectral lines and continuum fluxes, compared to 
the photometry flux in the V filter, F(V,$\lambda$5500\AA) shown in the top plot. Observations with different telescopes are denoted with different symbols shown in the upper two plots. The continuum flux is in units of 
$10^{-15} \rm erg \, cm^{-2} s^{-1}$\AA$^{-1}$ and the line flux in units of $10^{-13} \rm erg \, cm^{-2} s^{-1}$\AA$^{-1}$.} \label{fig2}
\end{figure*}

\begin{figure*}
\centering
\includegraphics[width=\columnwidth]{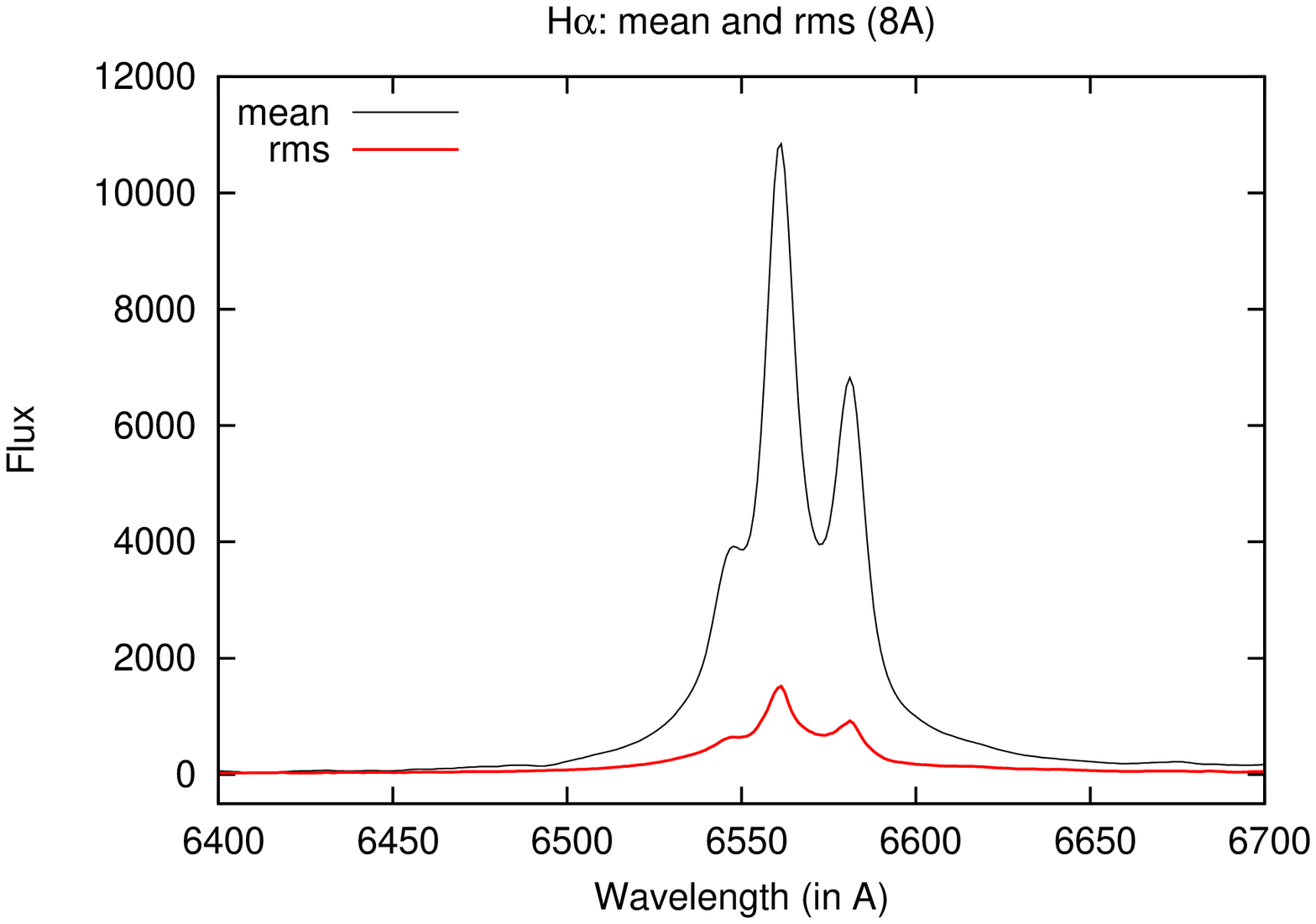}
\includegraphics[width=\columnwidth]{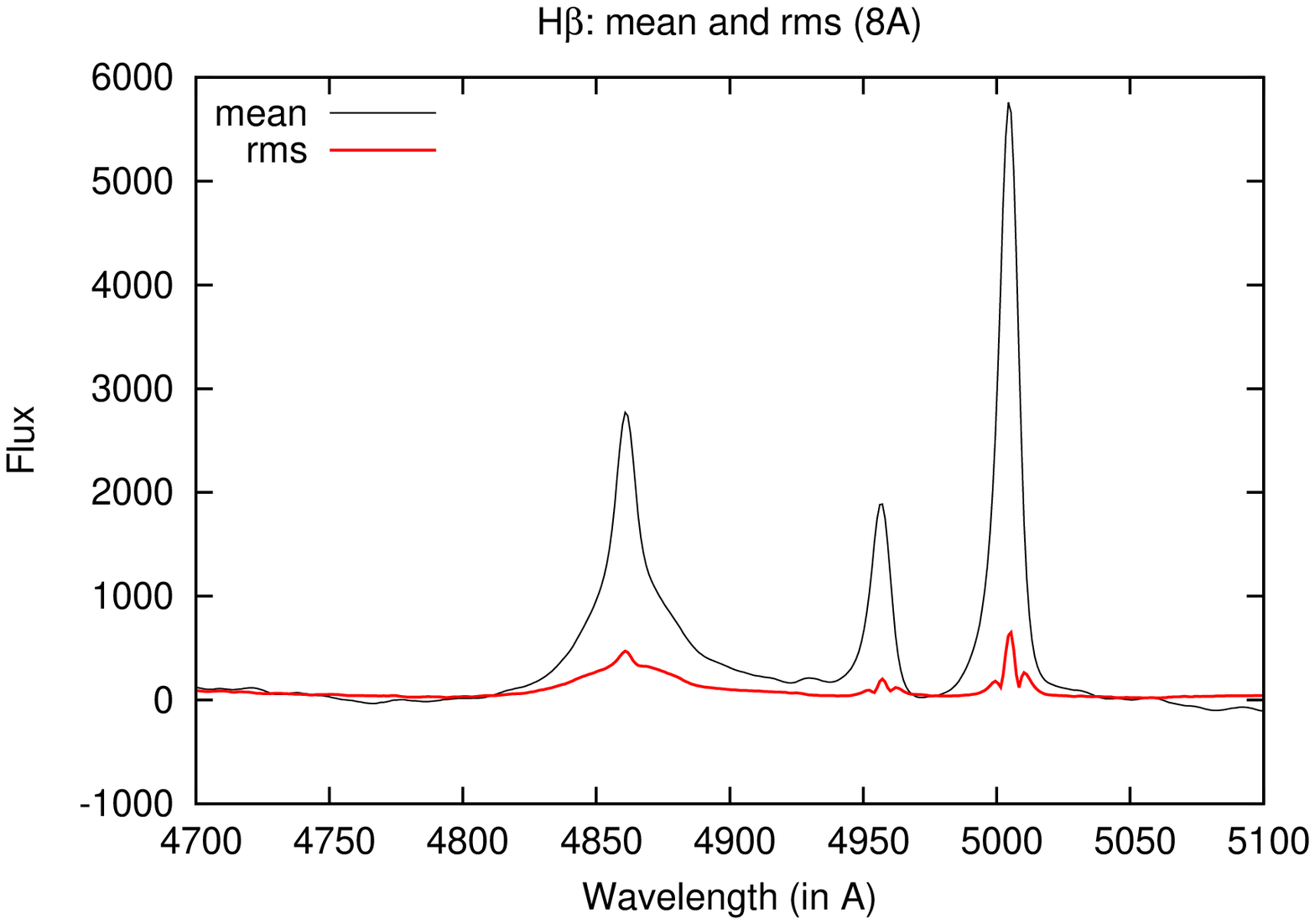}
\includegraphics[width=\columnwidth]{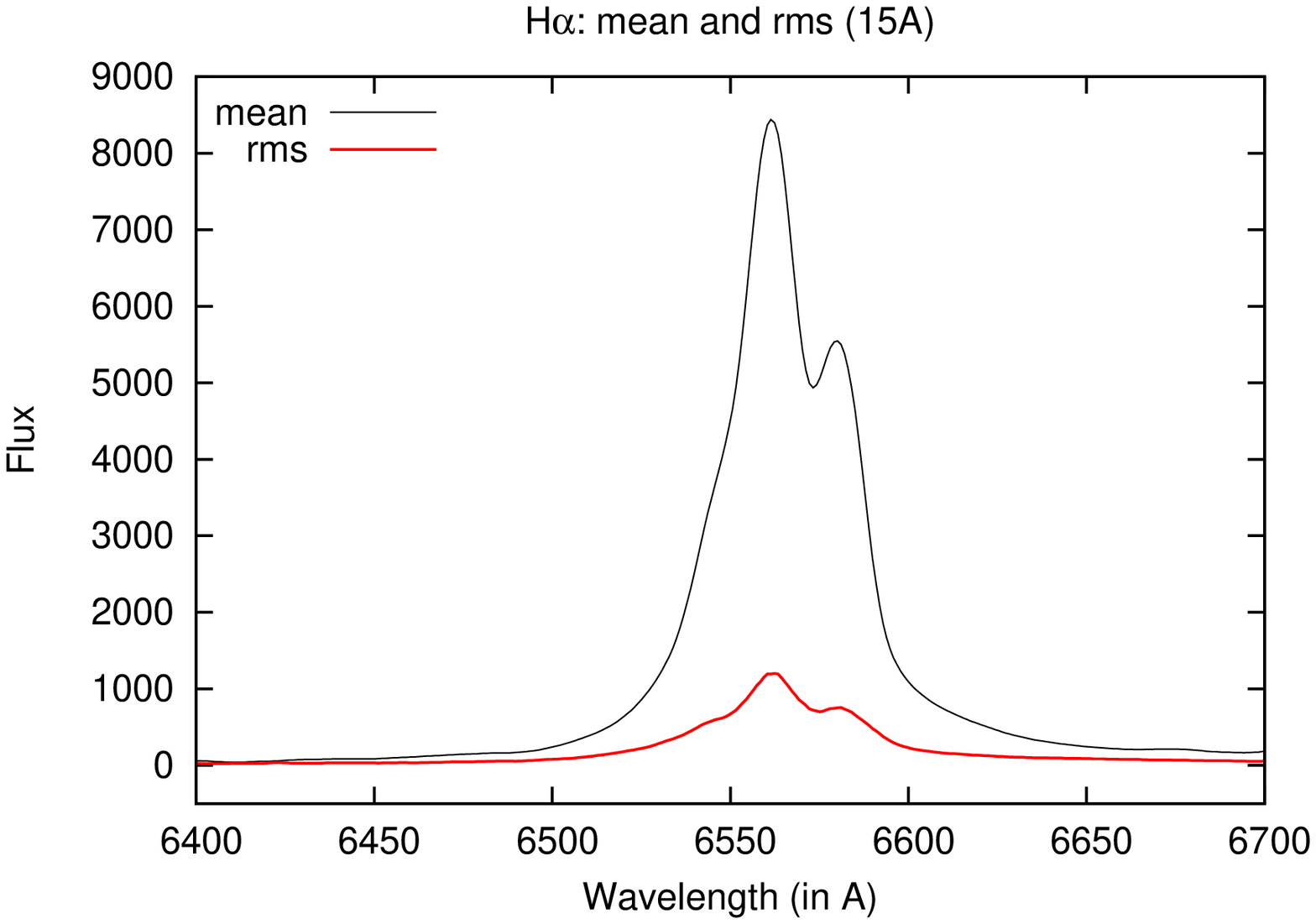}
\includegraphics[width=\columnwidth]{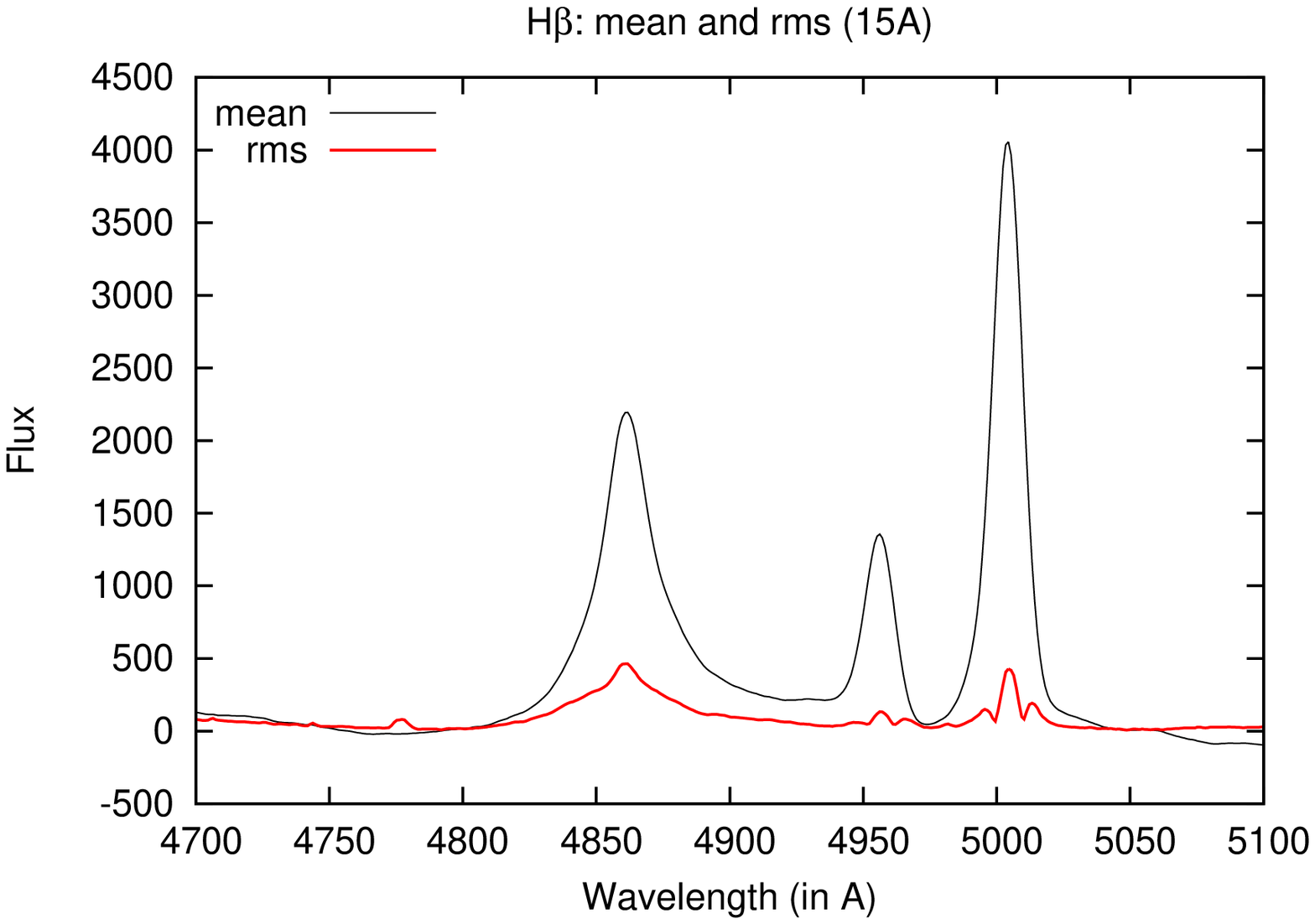}
\caption{Mean and rms profiles of the  H$\alpha$ (left) and H$\beta$ (right) lines using  spectra with 8\AA\, (upper panels) and 15\AA\, (bottom panels) spectral resolution.} \label{mean}
\end{figure*}

\section{Data analysis and results}

\subsection{Photometry}
\label{sec:photo}

The measured photometric data in the BVR filters for a circular aperture 15\arcsec and 
the corresponding errors are given in Table~\ref{tab1}, and shown in Fig.~\ref{fig1}.  As it
can be seen in Fig.~\ref{fig1}, the maximum of activity was observed in 1998.
Also there are several  noticeable outburst peaks  with different 
brightness amplitudes (Fig.~\ref{fig1}). For possible four outbursts in the V-photometric light-curve, 
we listed basic information in Table~\ref{flare-photo}. As can be seen  from  Table~\ref{flare-photo}, the outburst 
amplitude (i.e. the ratio of the maximal to  minimal flux - F(max/F(min)) varies from 1.1 to 1.6 times
in the intensity. The duration of outbursts (i.e. the times between two minimums) also differs.
The most prominent outburst peak (number 1 in Table~\ref{flare-photo}),
with the highest flux value and duration of $\sim$1766 days,  seen in Aug 19, 1998 (JD 51045.00).
The number 2 and 3 outbursts (durations of 727 and 989 days, respectively), 
may belong to the same outburst (2+3 in Table~\ref{flare-photo}) that has two peaks
in the intensity, which would have total duration of 1716 days.
The 4th outburst, with the duration of 1879 days, has 4 small peaks, the intensity of which varies 1.1--1.2
times relative to the neighbouring points. It is very interesting that the duration of
outburst number 1, (2+3), and 4 is in the range of  $\sim$(1700--1900) days (Table~\ref{flare-photo}), that
may indicate periodicity in the flux variability. 
Further (in Section~\ref{sec:period}) we study the possibility of a periodicity in the continuum and line light-curves.

\subsection{Spectral variability}

Fig.~2 shows the spectrum of NGC 7469, where the emission lines are denoted. Beside the 
Balmer emission lines there are the broad He I $\lambda$5876\AA\ and He II $\lambda$4686\AA , 
the broadband features from blended Fe II multiplets at $\sim$4670 \AA\ (Fe II 37 and 38 multiplets) and 5350 
\AA\ (Fe II 48 and 49 multiplets)\footnote{More details about the optical Fe II lines can be found in \citet{kov10} 
and \citet{sh12}, and also see http://servo.aob.rs/FeII\_AGN/ }. 
The broad lines are superposed with their narrow components and 
the forbidden narrow lines of  [O III]$\lambda\lambda$4959,5007\AA,  [O I]$\lambda\lambda$6300,6364\AA, 
[S II]$\lambda\lambda$6717,6731\AA, [Fe VII]$\lambda\lambda$5721,6087\AA\  etc.  
The common full width at half maximum (FWHM) of broad lines is $\sim$2000 km s$^{-1}$, and of narrow ones is 
$\sim$500km s$^{-1}$, that is typical for Seyfert 1 galaxies. 
Using the narrow line components we found an average redshift of z = 0.0164 which we accepted as cosmological one. It should be noted that the broad component of He II $\lambda$4686\AA\ line is very weak in  our spectra
showing a small flux (F(He II)$<2.0 \times 10^{-13}$) compared to Balmer broad lines, and its blue wing is superposed with Fe II$\lambda$4570\AA. In Fig.~\ref{fig-mm} we present the spectra of NGC 7469  in the minimum and maximum activity during the monitoring period. As can be seen in Fig.~\ref{fig-mm}, there is no big change in the spectral energy distribution between the minimum and maximum states. The continuum flux at 5100 \AA\
decreased by a factor of $\sim$2  (see fluxes in Table~\ref{tab5}) in 
the low-activity state, and the slope of the continuum in the blue range was 
flatter. Besides, the wings of the Balmer lines and the He II line became much weaker than in the high-activity state.

\subsubsection{Variability of the emission lines  and the optical continuum}
\label{sec:var}

We analysed flux variability of the emission lines and the optical continuum using 233 spectra  
of the  H$\beta$ line and continuum at 5100 \AA,  153 spectra  of the He II $\lambda$4686 line,  and 108 spectra of the H$\alpha$ line  and  red continuum at 6300 \AA. In Table~\ref{tab5}, continuum fluxes of the rest wavelength at 5100 \AA \, and  6300 \AA, and the  total H$\beta$, H$\alpha$,  and He II $\lambda$4686 lines are listed. In Fig.~\ref{fig2} the light-curves constructed from the data from Table~\ref{tab5} are presented for the H$\alpha$, H$\beta$ and He II $\lambda$4686  total line fluxes\footnote{We measured the total line fluxes where both the broad and narrow components are included.} and for the continuum at the rest  wavelength $\sim$5100 \AA \, and 6300 \AA.
For comparison, the first panel in Fig.~\ref{fig2}  shows the photometric light-curve in the V-filter. 
It can be clearly seen that the spectral variability is following the photometric one.
In the light-curves (see Fig.~\ref{fig2}) there are several flare-like peaks. 
These peaks  can be either due to the contribution of a ring having a diameter of  $\sim$5\arcsec($\sim$1.6 kpc) or  real flares (details are given  in Section~\ref{sec:flare}).

As usually \citep[see][]{sh08,sh12,sh13,sh16} we estimated  the variability rate
of the flux in the lines and in the continuum, using the method given by \citet{ob98}.
Final light-curve statistics for all five light-curves are given in Table~\ref{var}.
From Table~\ref{var} it can be seen that the  fluxes in the continuum (at 5100 \AA\ and 6300 \AA\ 
rest wavelengths) and the total  H$\alpha$ line flux changed for about 2  times, while  the  H$\beta$ 
flux was  changed  $\sim$3 times. The highest change is in  the He II $\lambda$4686\AA\ line, that is about $\sim$6 times.
The difference in the line flux variations between  the H$\alpha$, H$\beta$ and He II $\lambda$4686 lines
may be caused by different dimensions of the broad line emitting  regions.

We found that the amplitude of variability F(var) is $\sim$14\% for the continuum and total H$\alpha$ line,
and it is larger for the  H$\beta$ line ($\sim$23\%) and He II $\lambda$4686\AA\ ($\sim$32\%).

To explore the variability in the line profiles, we constructed the mean and root-mean-square
(rms) line profiles of the H$\alpha$ and H$\beta$ lines (see Fig.~\ref{mean}),
averaging separately the continuum  subtracted spectra with spectral resolution of $\sim$ 8 \AA\ and $\sim$ 15 \AA\ . 
Fig.~\ref{mean} shows that the broad line profiles (taking the both spectral resolutions) stay 
similar during the monitoring period\footnote{Note here that the narrow line residuals
in Fig.~\ref{mean} are due to slight difference in the spectral resolution.}. This indicates that the geometry of the line emission regions  remain  unchanged. 

\begin{figure*}
\centering
\includegraphics[width=\columnwidth]{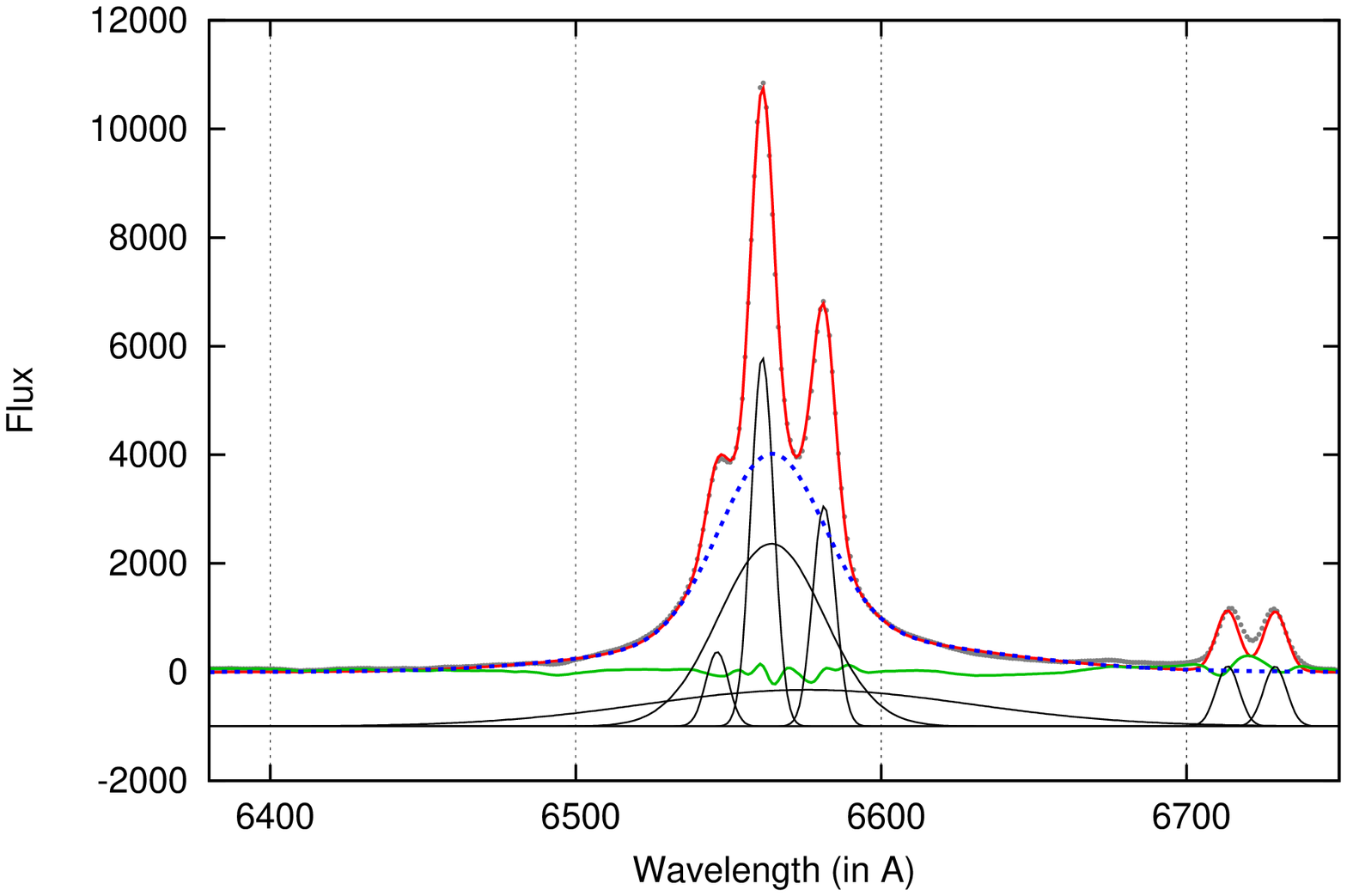}
\includegraphics[width=\columnwidth]{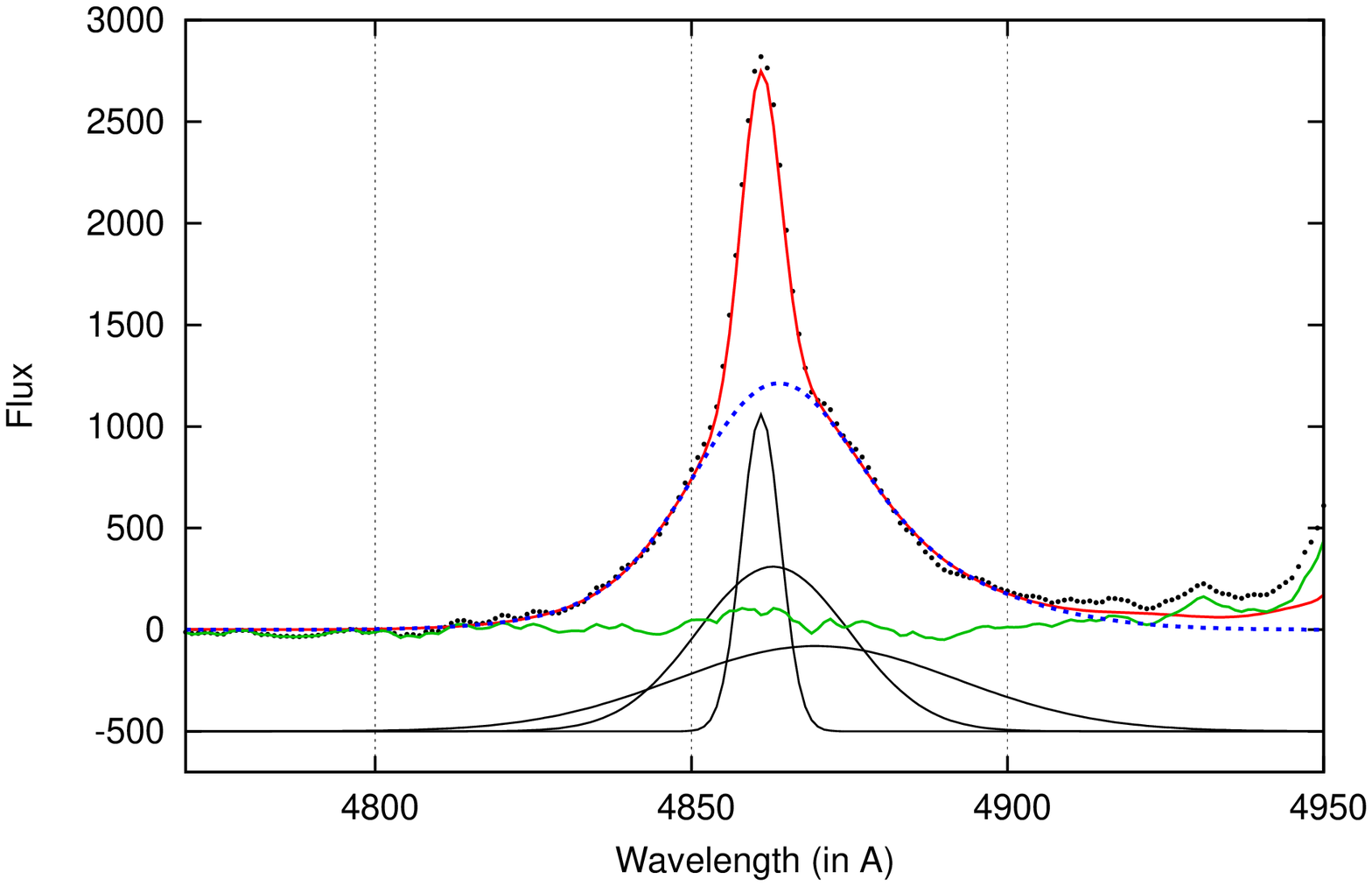}
\caption{The Gauss best-fitting of the mean H$\alpha$ (left) and H$\beta$ (right) lines (dots) of the 8\AA\, spectral resolution (the same as in Fig.~\ref{mean}, upper panels). The model is given with solid line, the modeled broad line component with dashed line, and the residual is shown underneath. The Gaussian components are shown shifted below (solid lines).} \label{broad}
\end{figure*}

To measure the FWHM of the broad H$\alpha$ and H$\beta$ line components, we fitted the mean profiles of the 
lines with a number of Gaussian functions using the $\chi^2$ minimization \citep[see, e.g.][]{pop04}  with the aim to
subtract the contribution of all narrow lines to the mean profiles. As it can be seen in Fig.~\ref{broad}, 
it seems that broad components of H$\alpha$ and H$\beta$ are more complex that can be presented by only one broad Gaussian. 
There is a very broad component shifted to the red ($\sim 600$ km s$^{-1}$).
The slightly red asymmetry is present in H$\alpha$ and H$\beta$ during the whole monitoring period.
Taking only the broad component of the mean profiles (represented with two broad Gaussian functions - 
central and one very broad, slightly redshifted, see Fig.~\ref{broad}) we found that the FWHM of 
the broad H$\alpha$ component is $\sim$2100 km s$^{-1}$ that seems to be slightly larger than the FWHM of
the broad H$\beta$ component ($\sim$2000 km s$^{-1}$).

\begin{table}
\centering
\caption{Outbursts detected  in the V-photometric light-curve of NGC 7469. 
Columns are: (1): Number of flare-like event. (2): UT-date. (3): Modified Julian Date. (4): The V-flux in 
units $10^{-15} \rm erg \ cm^{-2} s^{-1} \AA^{-1}$. (5): Ratio of the maximal to minimal flux. 
(6): Difference in days between adjacent minima.} 
\label{flare-photo}
\begin{tabular}{lccccc}
\hline
N & UT-date & JD+ & F(V) & F(max)/F(min)& $\Delta$ t \\
 & & 2400000 &  & & day\\
1 & 2 & 3 & 4 & 5 & 6 \\
\hline
 1  & 19.08.1998  &  51045.00  &   25.12   &  1.55  &     1766     \\
\hline
 2  & 30.08.2005  &  53613.49  &   20.00  &   1.38   &    727    \\ 
 3  & 06.09.2007  &  54349.47  &   18.20  &   1.20   &    989    \\ 
\hline
2+3 &            &            &           &         &    1716 \\
\hline
 4  & 27.09.2010  &  55467.32  &   20.42   &  1.44   &    1979   \\  
\hline
\end{tabular}
\end{table}

\begin{table}
\centering
\caption{Parameters of the continuum and line variations. Columns are: (1): Analysed feature of the spectrum. (2): Total number of spectra. (3): Mean flux. (4): Standard
deviation. (5): Ratio of the maximal to minimal flux. (6): Variation amplitude (see text).
Continuum flux is in units of $10^{-15} \rm erg \ cm^{-2} s^{-1} \AA^{-1}$
and line flux in units of $10^{-13} \rm erg \ cm^{-2}s^{-1}$. }
\label{var}
\begin{tabular}{lccccc}
Feature & N  & $F$(mean) &  $\sigma$($F$) & $R$(max/min)& $F$(var)\\
1 & 2 & 3 & 4 & 5 & 6  \\
\hline
 cont 6300         & 108 &    14.64   & 2.36   &  2.21    &  0.158 \\
 cont 5100         & 233 &    16.78   & 2.55   &  2.13    &  0.149 \\
 H$\alpha$ - total & 108 &    37.26   & 5.42   &  2.01    &  0.144 \\
 H$\beta$ - total  & 233 &     7.99   & 1.89   &  3.04    &  0.233 \\
 He II - total     & 153 &     2.92   & 0.94   &  5.75    &  0.313 \\ 
\hline
\end{tabular}
\end{table}


\subsubsection{Flare-like peaks}
\label{sec:flare}

As noted  above  (see Section~\ref{sec:var}), the light-curves exhibit few flare-like peaks, in which the flux increases for about 10\% -- 30\%  in several days.  These peaks  can be caused both by the contribution from the circumnuclear star-forming ring with the diameter of  $\sim$5\arcsec\ 
(Section~\ref{sec:star}) and by the processes in the AGN which produce flare or flare-like events.
In Section~\ref{sec:star} we discussed that  the movement of the slit can contribute to the artificial 
flare-like peaks, and consequently we excluded spectra where this was the case. However, in 
the light-curves we can still see the short flare-like peaks.  Unlike the long lasting outbursts observed in the photometric light-curves (of the order of hundreds to thousands days), these events are lasting significantly shorter, of around several (1--5) days.
We repeat that we estimated the mean uncertainties in the line and continuum fluxes using observational data, 
separated by 1--3 days (see Section~\ref{sec:flux}), for which we have around $\sim$5\% cases with large difference in the flux ($\sim$10\%-30\%), indicating flare-like events. As noted before, we accounted the spectroscopic flare-like events, only if we could find the counterpart in the photometric light-curve.

In Table~\ref{flare-spec} we list some information for the possible flare-like events detected in the spectral light-curves. From Table~\ref{flare-spec}, it can be seen that during the flares, which is lasting between 1 and 3 days, the flux amplitudes vary within 1.1--1.4 times. 
The most prominent flares, which are seen in the photometric and spectroscopic  light-curves are noted as 
2 and 3 in Table~\ref{flare-photo} and denoted as the spectral flares 4 and 5 in Table~\ref{flare-spec}.

\begin{figure*}
\centering
\includegraphics[width=\columnwidth]{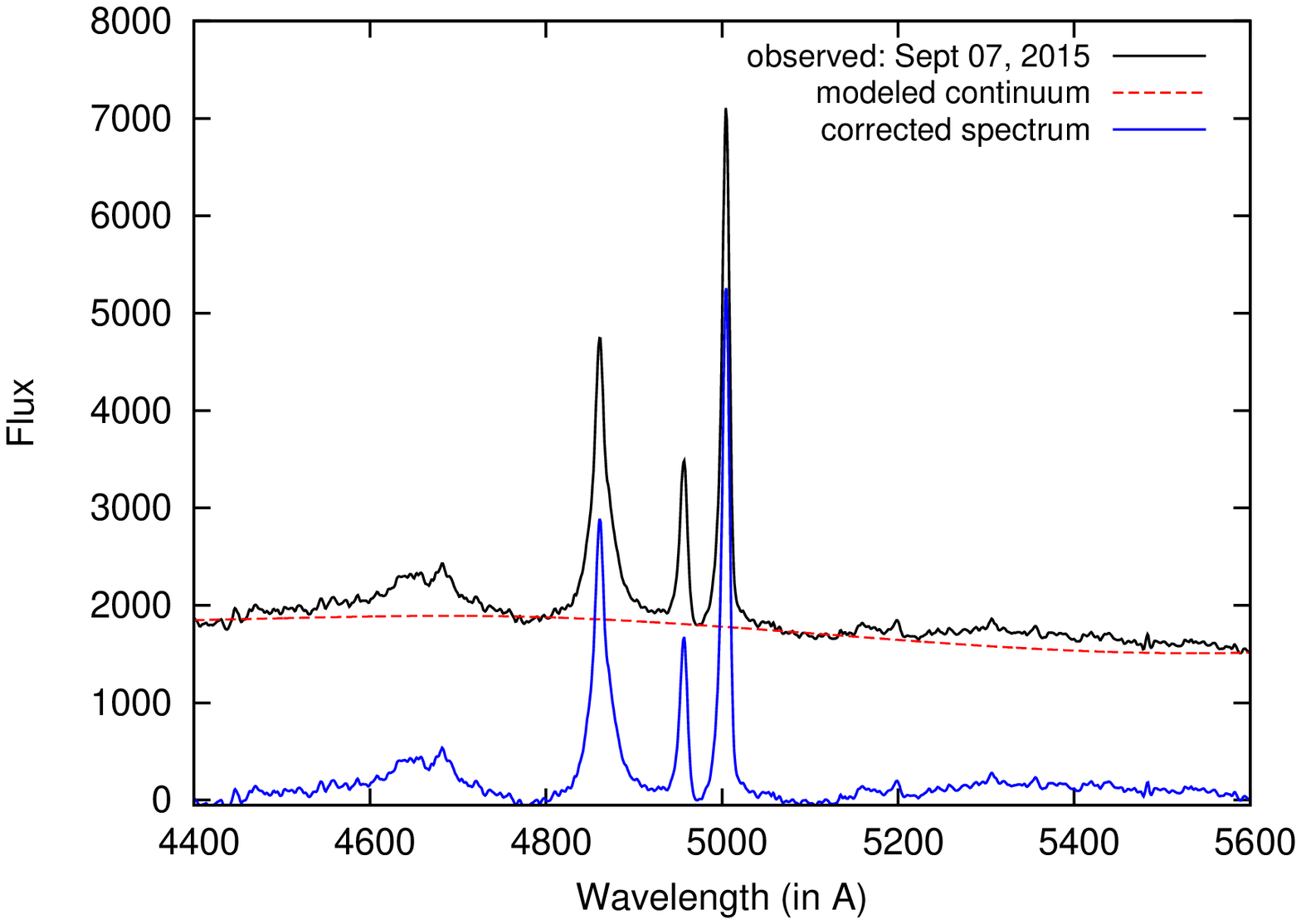}
\includegraphics[width=\columnwidth]{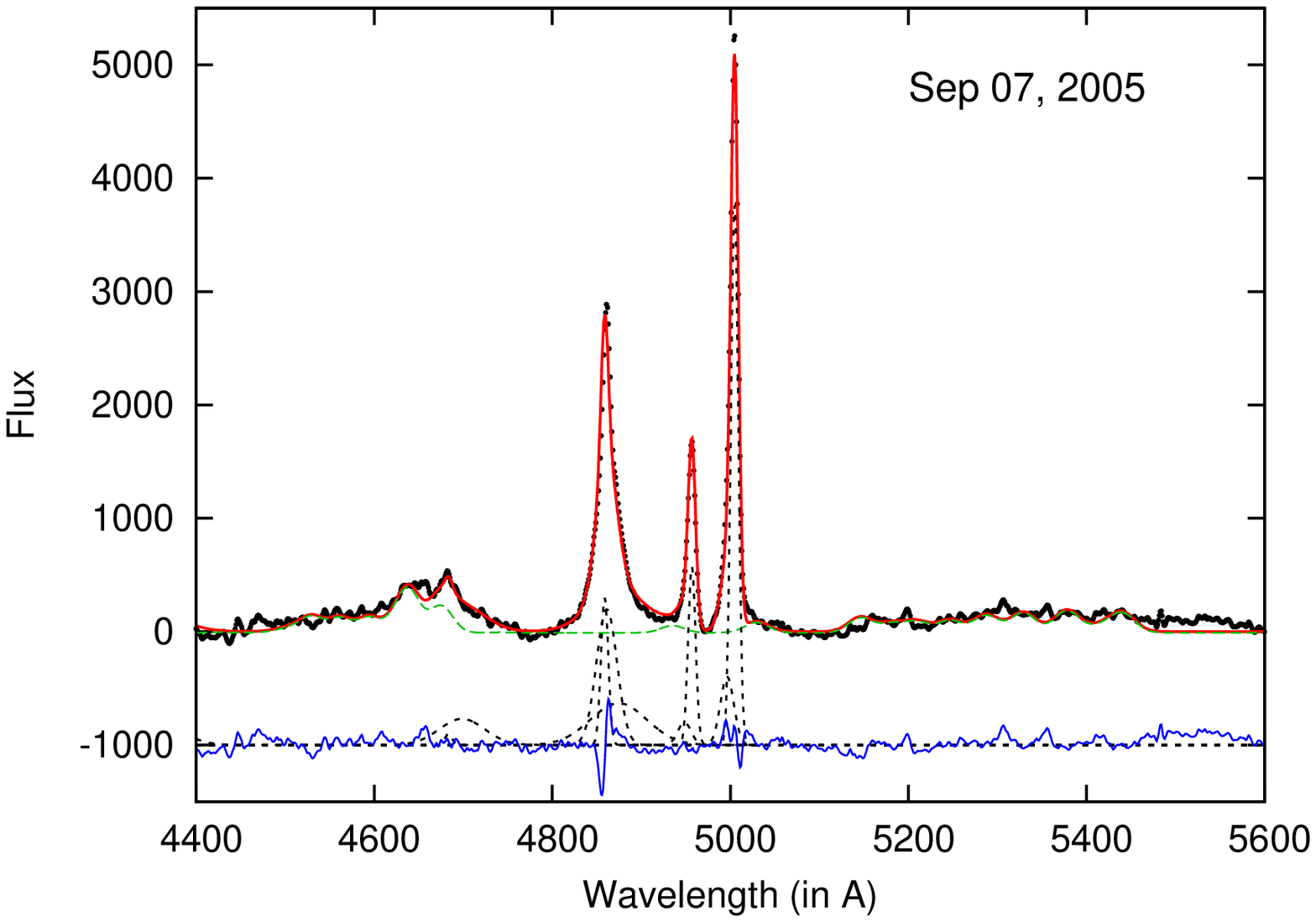}
\includegraphics[width=\columnwidth]{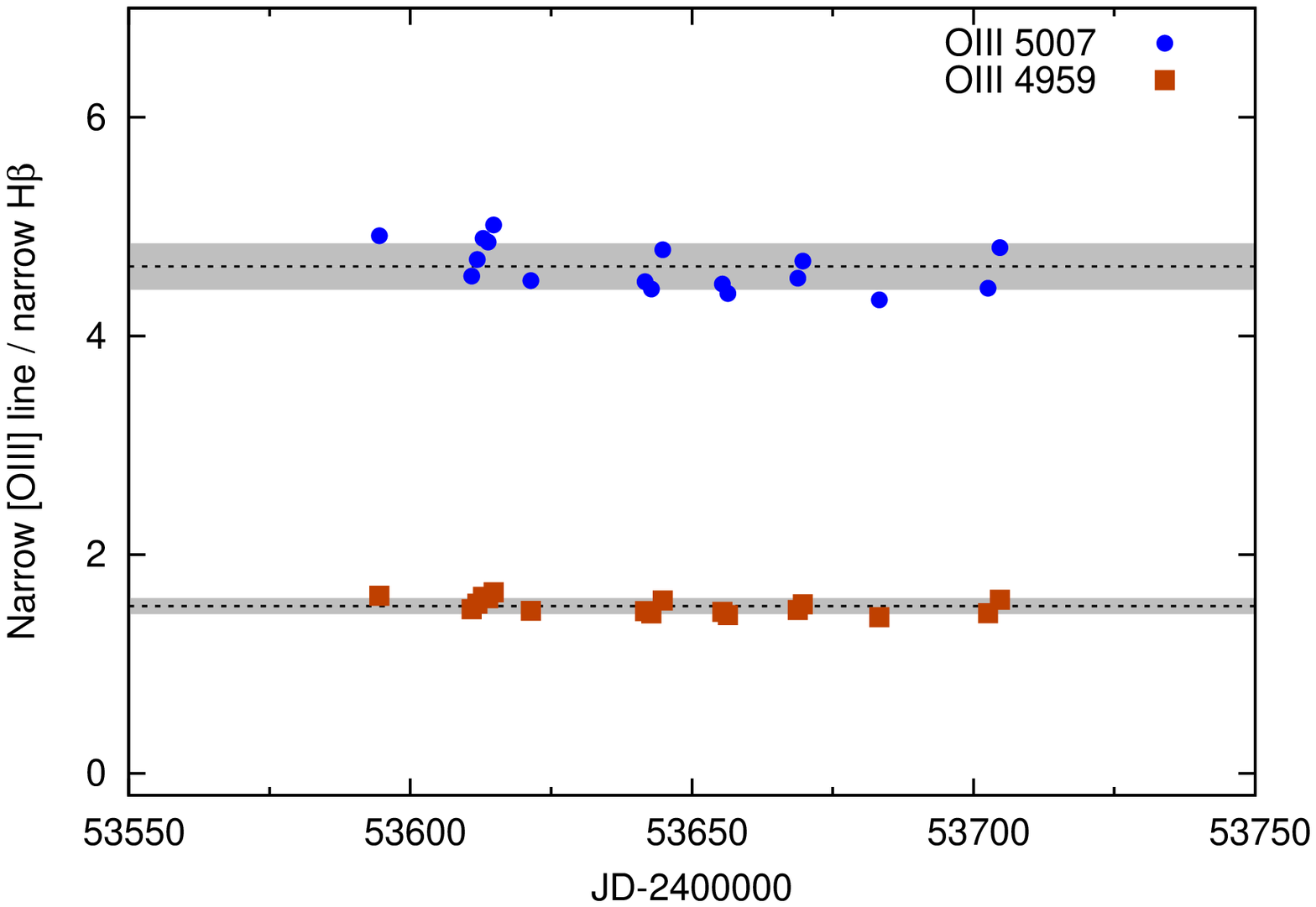}
\includegraphics[width=\columnwidth]{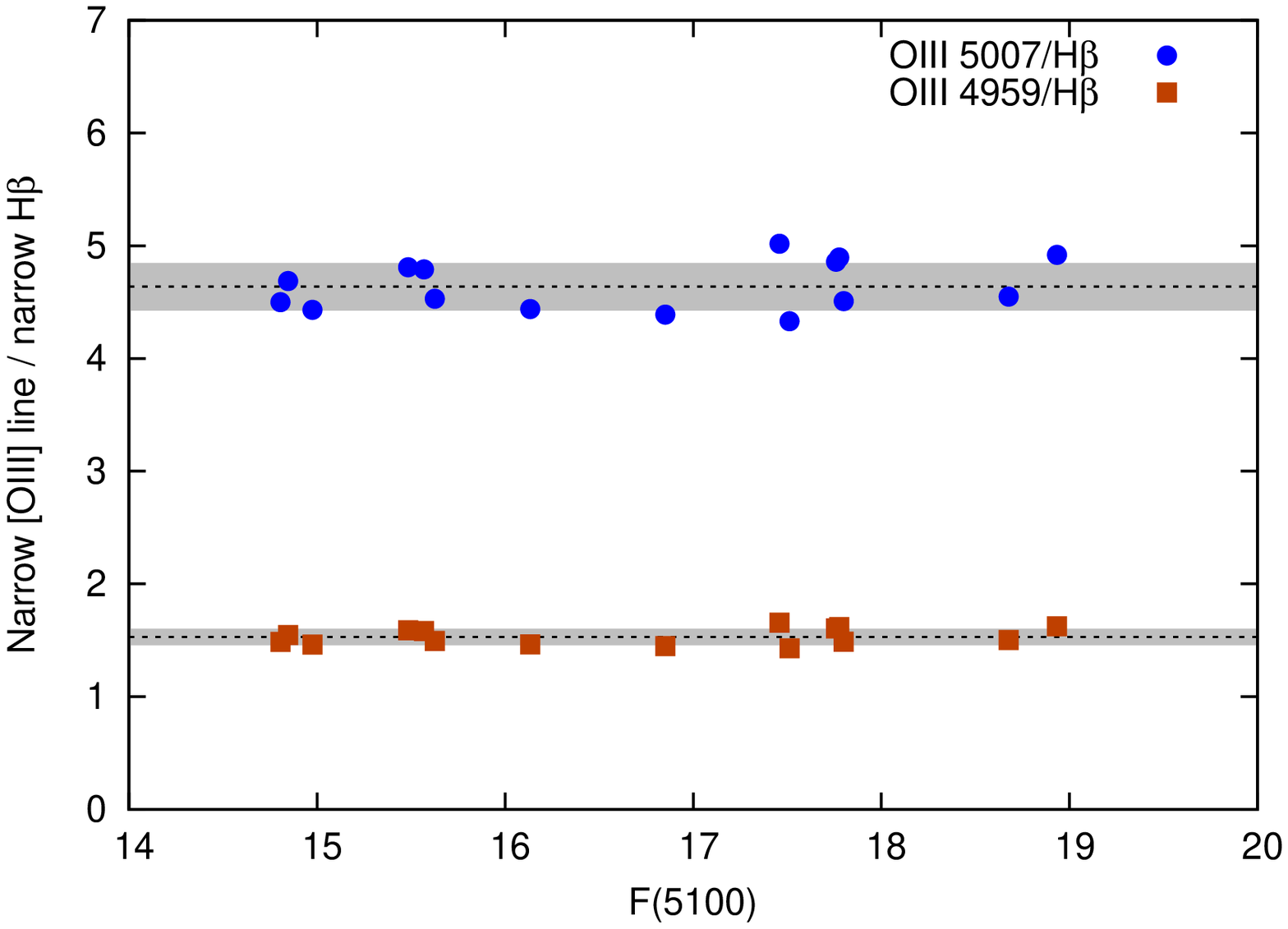}
\caption{Example of the continuum subtraction  using cubic spline (upper 
left) and multi-Gaussian best-fitting (upper right) of the spectral range 4200--5600\AA.
Below are shown the change of the narrow lines to H$\beta$ narrow line flux ratio during 2005 flare-like event (left) and the narrow-line to H$\beta$ (narrow) line flux ratio vs. continuum flux at 5100 \AA\, (right).} \label{fit}
\end{figure*}

Additionally, we examine the measurements of the ratio of the narrow lines that should systematically change in the case of the slit offset effect. { For this we first subtracted the underlying continuum (Fig.~\ref{fit}, upper left panel), and then performed the multi-Gaussian best-fitting described in \citet{pop04}, which includes the Fe II template from \citet{kov10,sh12} (see Fig.~\ref{fit}, upper right panel), and finally extracted the narrow lines.} 
For the flare-like peak observed in 2005, we found the flux of the narrow lines only, 
and then we explore if there is a change in the narrow line ratios (Fig.~\ref{fit}, bottom left panel) and in the narrow line ratio as a function of the continuum at 5100 \AA\ (Fig.~\ref{fit}, bottom right panel). Fig.~\ref{fit} illustrates that there is no trend in the ratio of the narrow lines, which is expected in the slit offset effect. 
It seems that the flare-like peaks in the NGC 7469 light-curve are real. The nature of the peaks may be that we have some short time outburst which have been detected in other AGNs \citep[see e.g.][]{sh12}. Additionally, in the photometric light-curves given in \cite{dor10}, the presence of flare-like peaks is obvious (see Fig.~1 in their paper).

\begin{table*}
\centering
\caption{Possible flares in spectral light-curves of NGC 7469. The columns are: (1): Number of spectral flare-like event. (2): UT-date, (3): Modified Julian Date, (4): Difference in days between
the event with the maximal flux and the nearest point (event) on the light-curve, 
(5-9): Ratio of the maximal flux in the flare to the flux of the nearest
point - for continuum at 5100 \AA, H$\beta$, He II 4686 \AA, continuum at 6300 \AA, and H$\alpha$, 
respectively, (10): Modified Julian Date for maximal photometric flare, and
(11): Number of photometric flare from Table~\ref{flare-photo}.}
\label{flare-spec}
\begin{tabular}{lcccccccccc}
\hline
N & UT-date & JD+ & $\Delta$ t & \multicolumn{5}{c}{F(max)/F(nearest)} & JD+ & N \\
spec. &  & 2400000 & day  & 5100 & H$\beta$ & He II& 6300 & H$\alpha$ & 240000 & \\
      &  & spec.   &      &      &          &      &      &           & phot.  & \\
1 & 2 & 3 & 4 & 5 & 6 & 7 & 8 & 9 & 10 & 11 \\
\hline
 1& 25.10.1998 &51083.73 & 1      & 1.26 & 1.13 & 1.24 & 1.26 & 1.05 &  51045 &   1  \\                                       
 2& 24.07.2001 &52115.47 & 3      & 1.37 & 1.15 & 1.05 & -    &  -   &  52107 &      \\          
 3& 06.09.2004 &53254.95 & 1      & 1.20 & 1.04 & 1.20 & 1.00 & 1.16 &  53242 &      \\                                    
 4& 11.10.2005 &53655.14 & 1      & 1.25 & 1.20 & -    & 1.21 & 1.07 &  53613 &   2  \\               
 5& 17.10.2007 &54391.39 & 2      & 1.08 & 1.01 & 1.22 & 1.08 & 1.25 &  54349 &   3  \\                 
 6& 12.11.2010 &55513.38 &$\sim$75 &1.17 & 1.25 & 1.22 & 1.21 & 1.22 &  55467 &   4  \\
\hline
\end{tabular}
\end{table*}

\begin{figure*}
\centering
\includegraphics[width=\columnwidth]{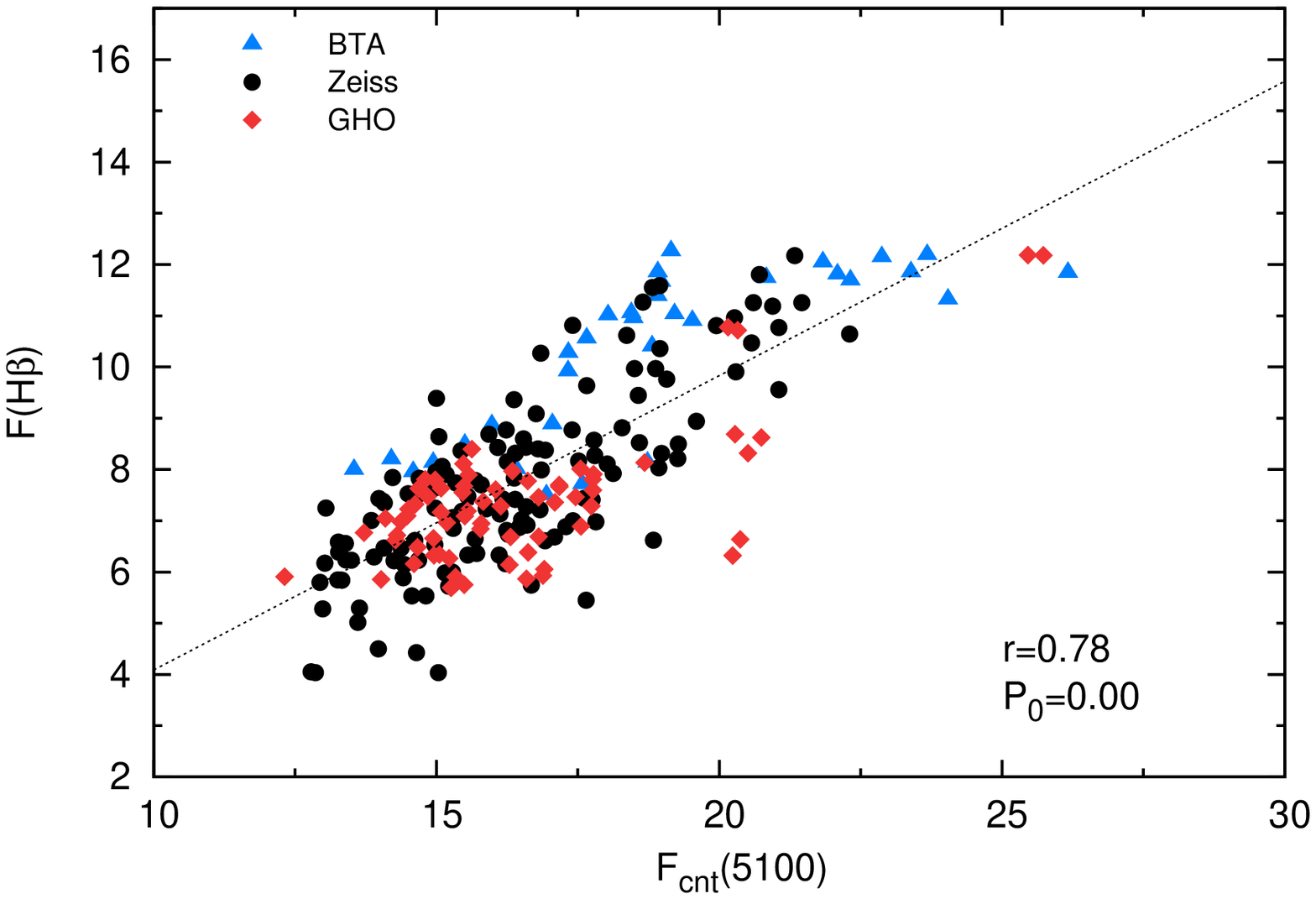}
\includegraphics[width=\columnwidth]{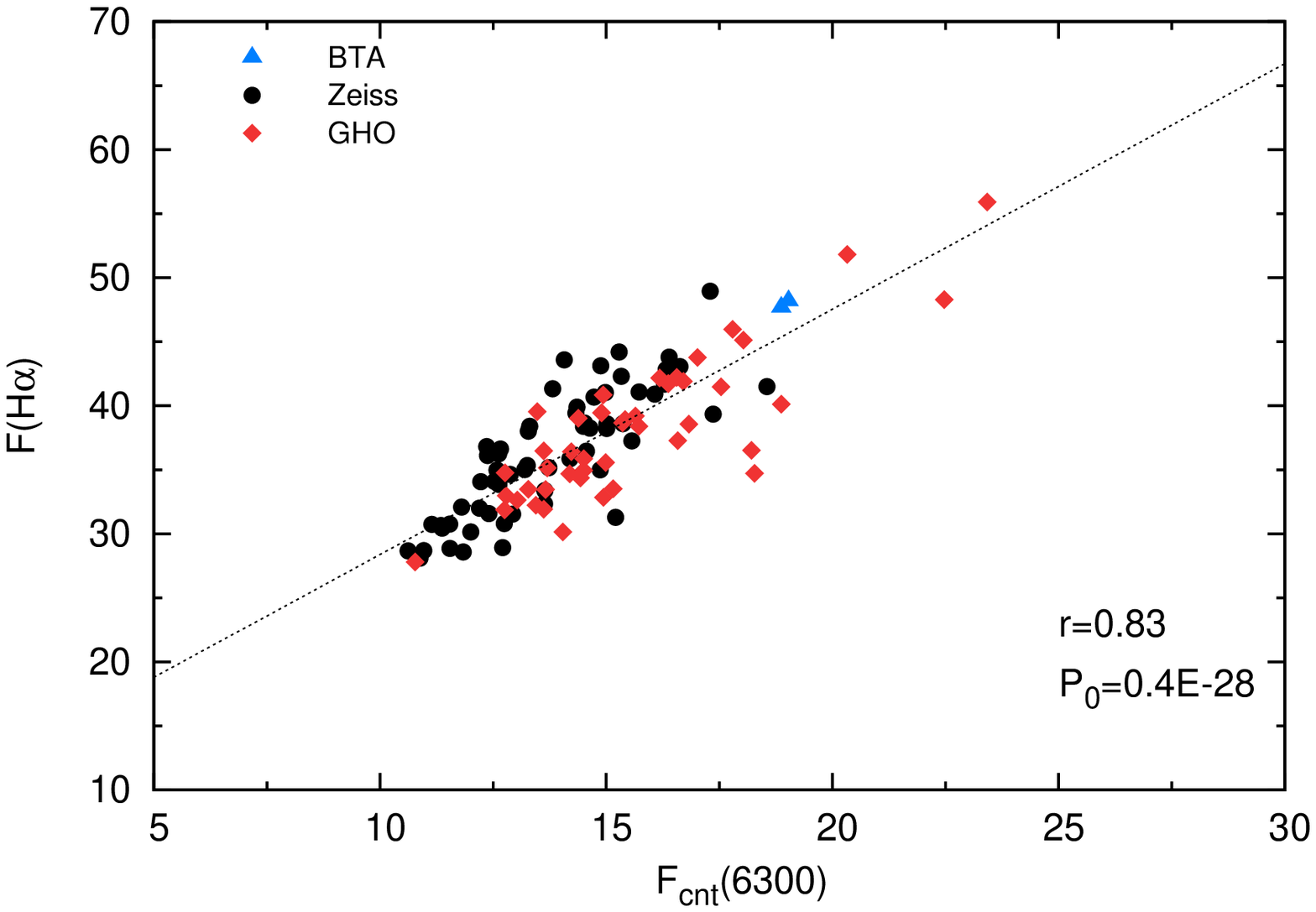}
\caption{Continuum vs. line flux for H$\beta$ (left) and H$\alpha$ (right). Symbols and units are the same as in Fig.\ref{fig2}.} \label{fig3}
\end{figure*}

\begin{figure*}
\centering
\includegraphics[width=\columnwidth]{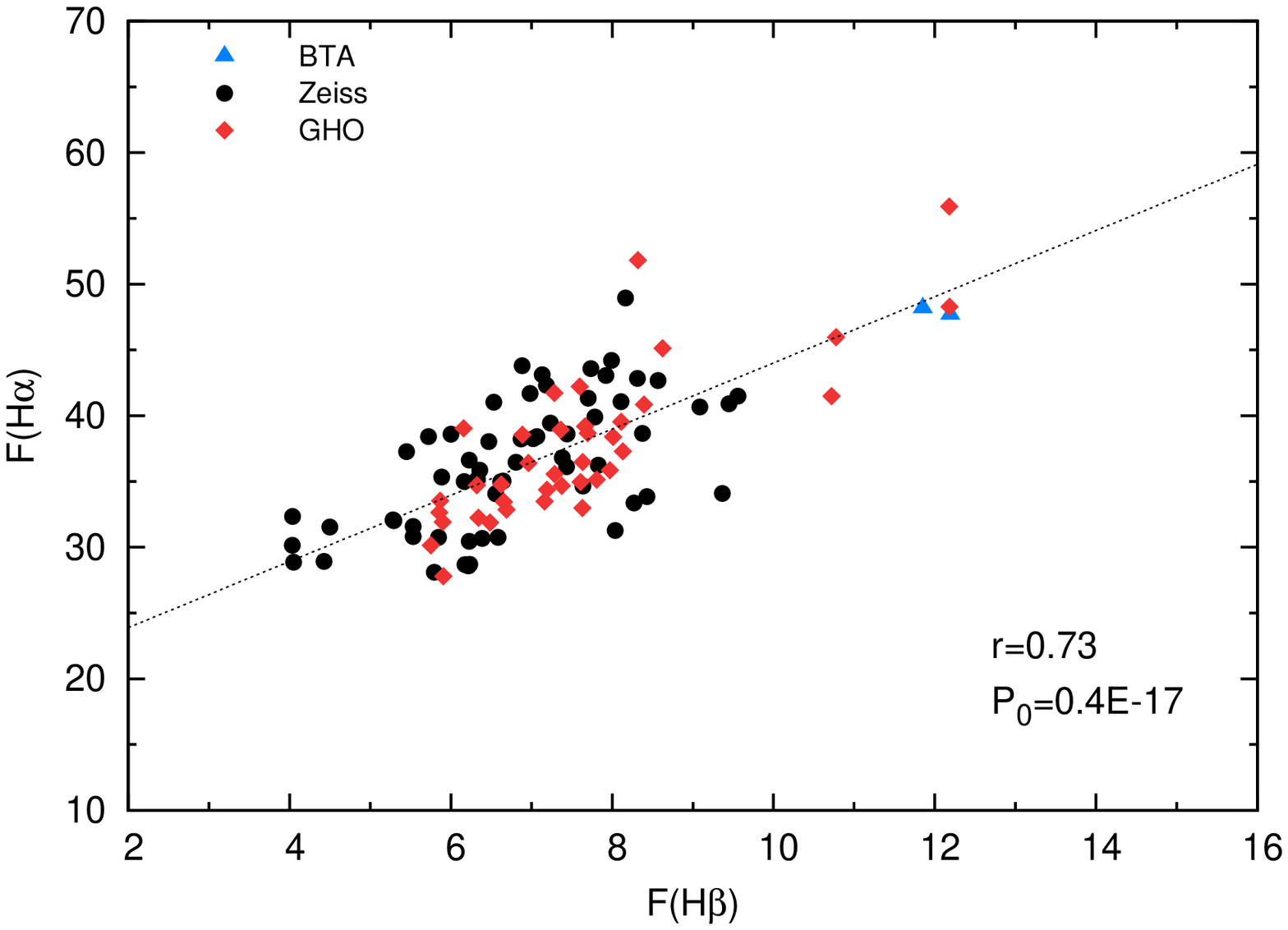}
\includegraphics[width=\columnwidth]{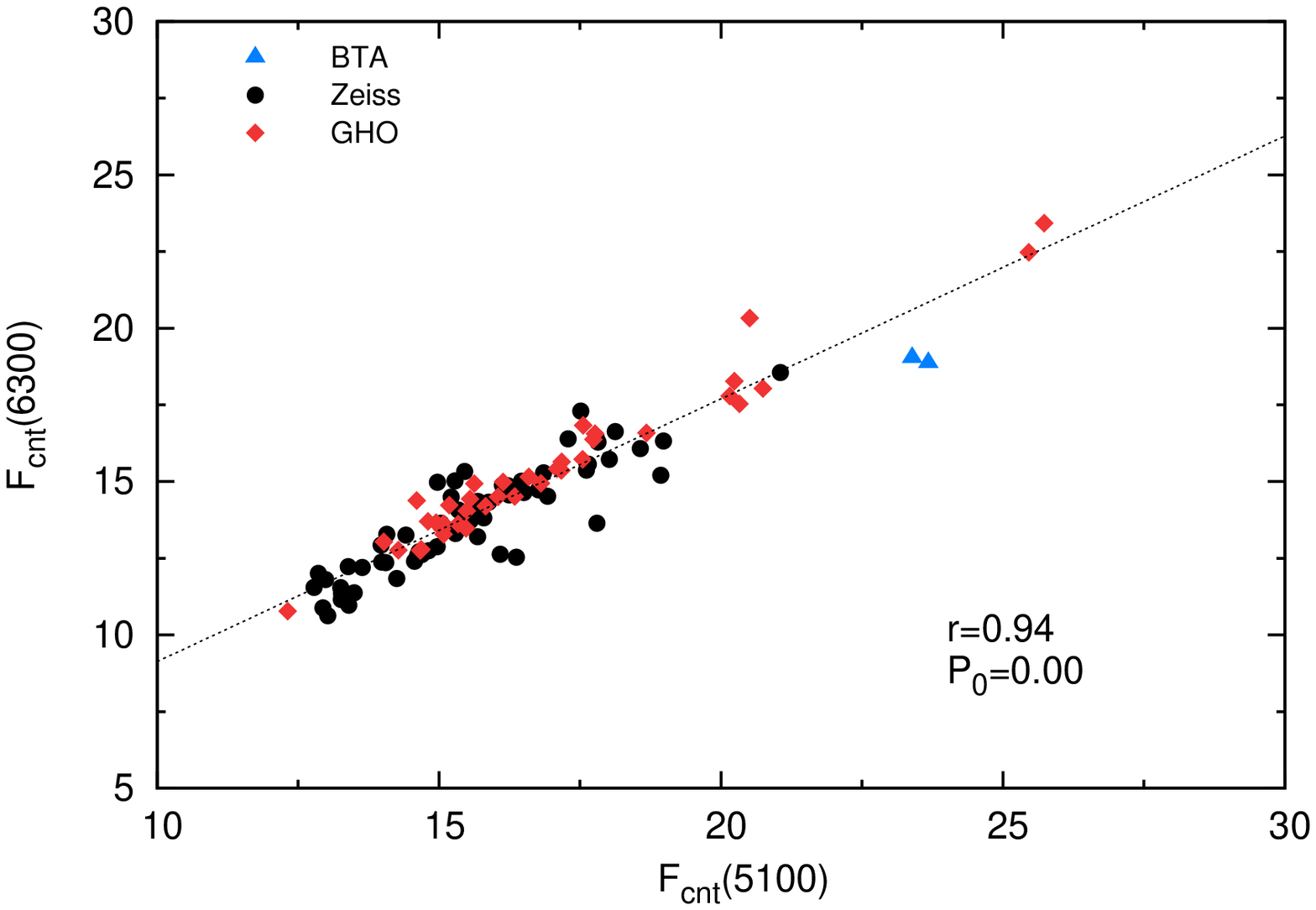}
\caption{H$\alpha$ vs. H$\beta$ line flux (left) and red vs. blue continuum flux (right). Symbols and units are the same as in Fig.\ref{fig2}.} \label{fig4}
\end{figure*}

\begin{figure*}
\centering
\includegraphics[width=\columnwidth]{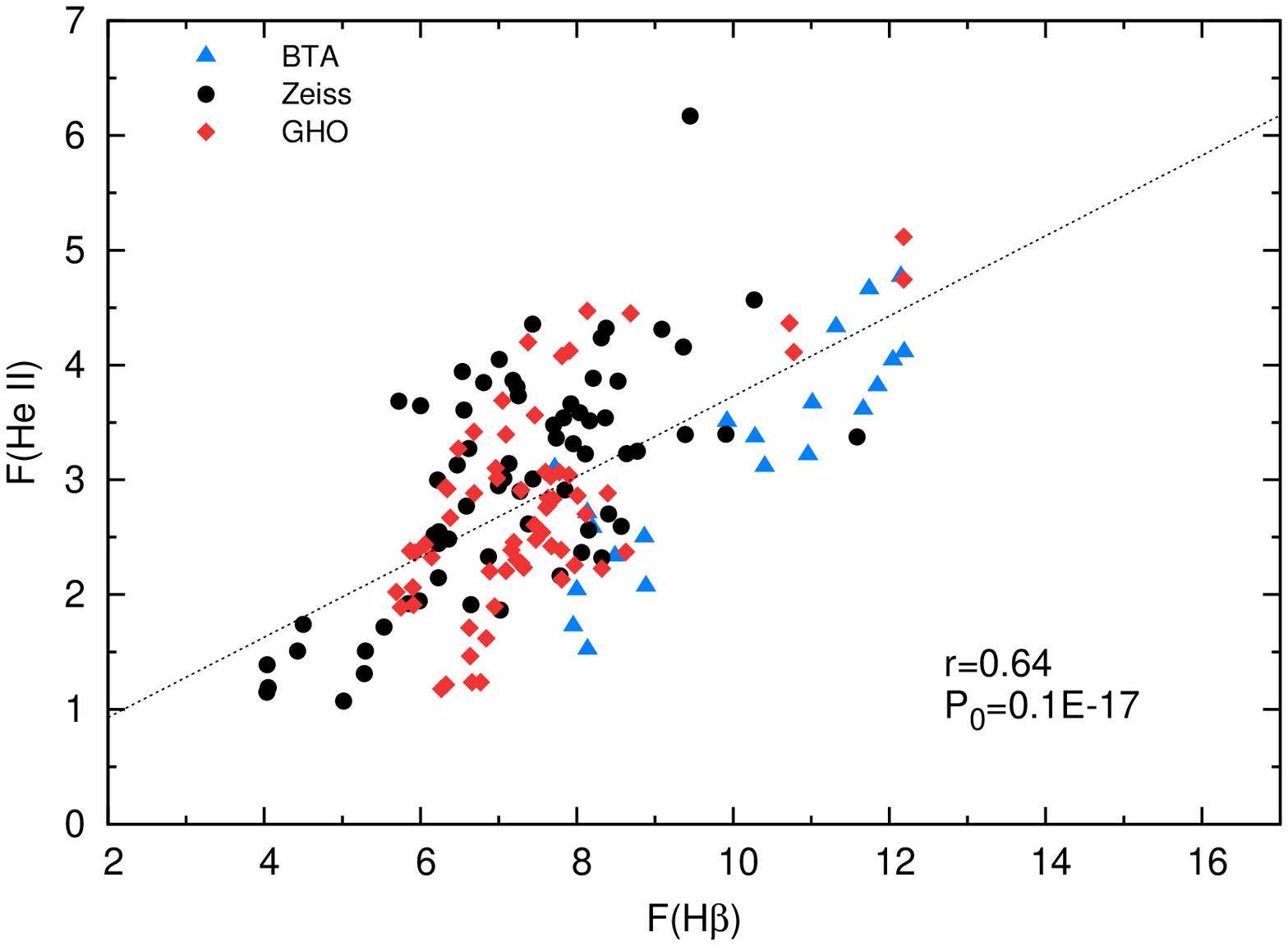}
\includegraphics[width=\columnwidth]{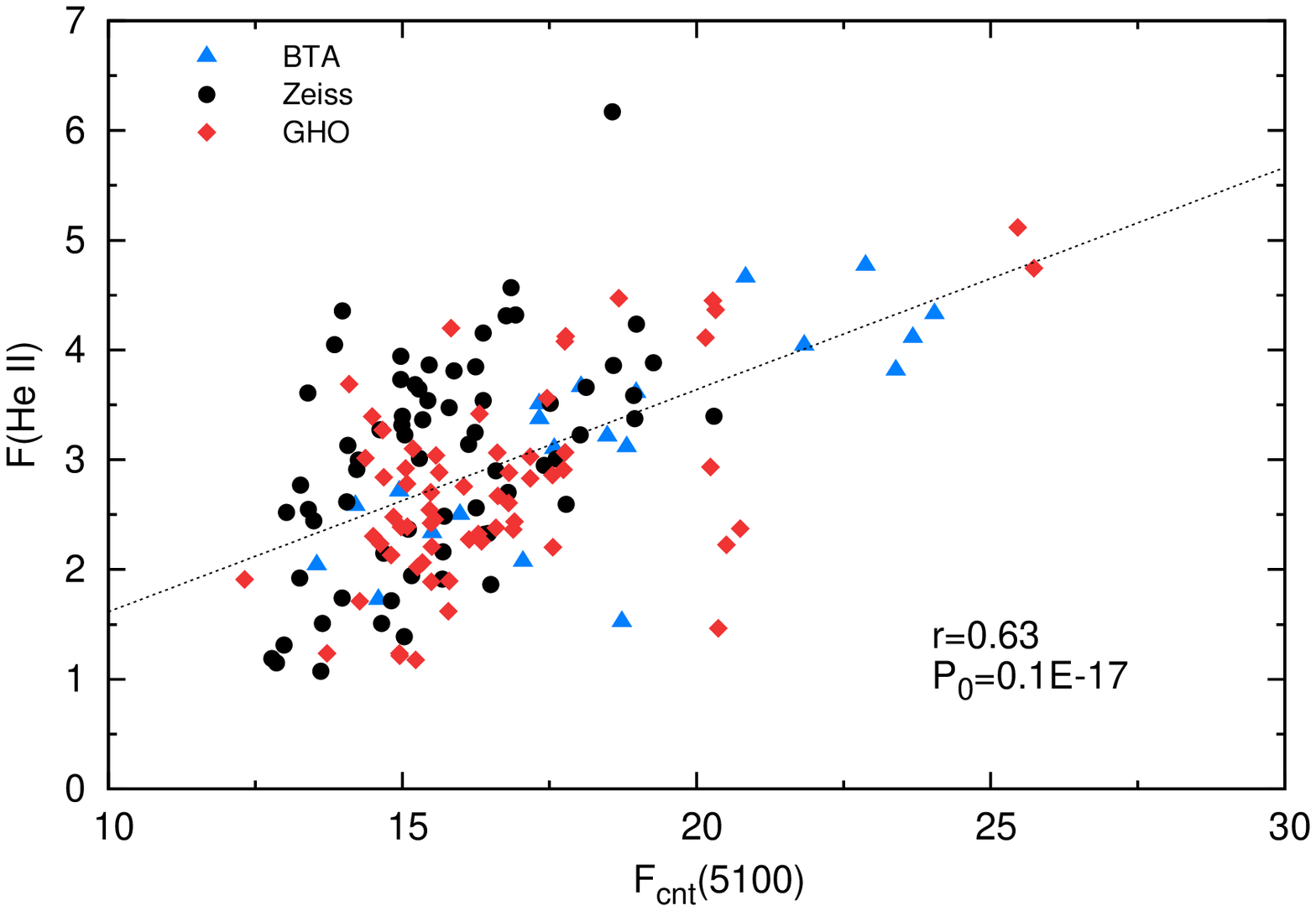}
\caption{HeII vs. H$\beta$ line flux (left) and blue continuum flux (right). Symbols and units are the same as in Fig.\ref{fig2}.} \label{fig5}
\end{figure*}

\subsubsection{Line and continuum flux correlations}
\label{sec:correl}

To explore the correlation between lines and corresponding continuum,  in Fig.~\ref{fig3} we plot the fluxes of 
H$\beta$ vs. 5100 \AA\, continuum and H$\alpha$ vs. 6300 \AA\, continuum. 
Fig.~\ref{fig3} shows that there are  good correlations between the line fluxes of H$\beta$/H$\alpha$ and the
corresponding continuum. 
The correlation coefficients  in both cases are quite high (r$\sim$0.8), indicating that
 the ionizing continuum is a good extrapolation of the optical one. 
 This is supported by the fact that the correlation between the blue 
 ($\sim$5100\AA) and the red ($\sim$6300\AA) continua is very high (r$\sim$0.94, Fig.~\ref{fig4}).
However, the correlation between  the flux of He II $\lambda$4686\AA\  with the continuum 5100 \AA\, and H$\beta$ is very low (see Fig.~\ref{fig5}). This is probably coming from the uncertainties of the He II flux measurements, since the line is very weak, especially in the low state of the AGN activity. { Moreover, the He II line is influenced with the Fe II multiplets (see dashed line underneath the observed spectrum in Fig.~\ref{fit}, upper right).}
Higher scatter of points is seen for lower values of the He II flux (right panel in Fig.~\ref{fig5}).
Also the correlations between H$\alpha$ vs. H$\beta$ is slightly smaller
(r$\sim$0.73, Fig.\ref{fig4}) than their correlation with the continuum.
The latter may be due to the fact that, as we noted above (Section~\ref{sec:flux}), the H$\alpha$ line can be affected by very small slit position offset.

\subsubsection{Time-lag analysis}
\label{sec:lag}

We applied the  ZDCF and SPEAR statistical methods to calculate the time
delays of continuum and line light-curves. Detailed description, examination,
and comparison  of these methods is given in \citet{ko14}.
Notice that one of  the employed methods (SPEAR) is also used by \citet{pet14}
for the same object, but for a set of spectra observed in a  much shorter period.

\begin{figure*}
\centering
\includegraphics[width=5cm]{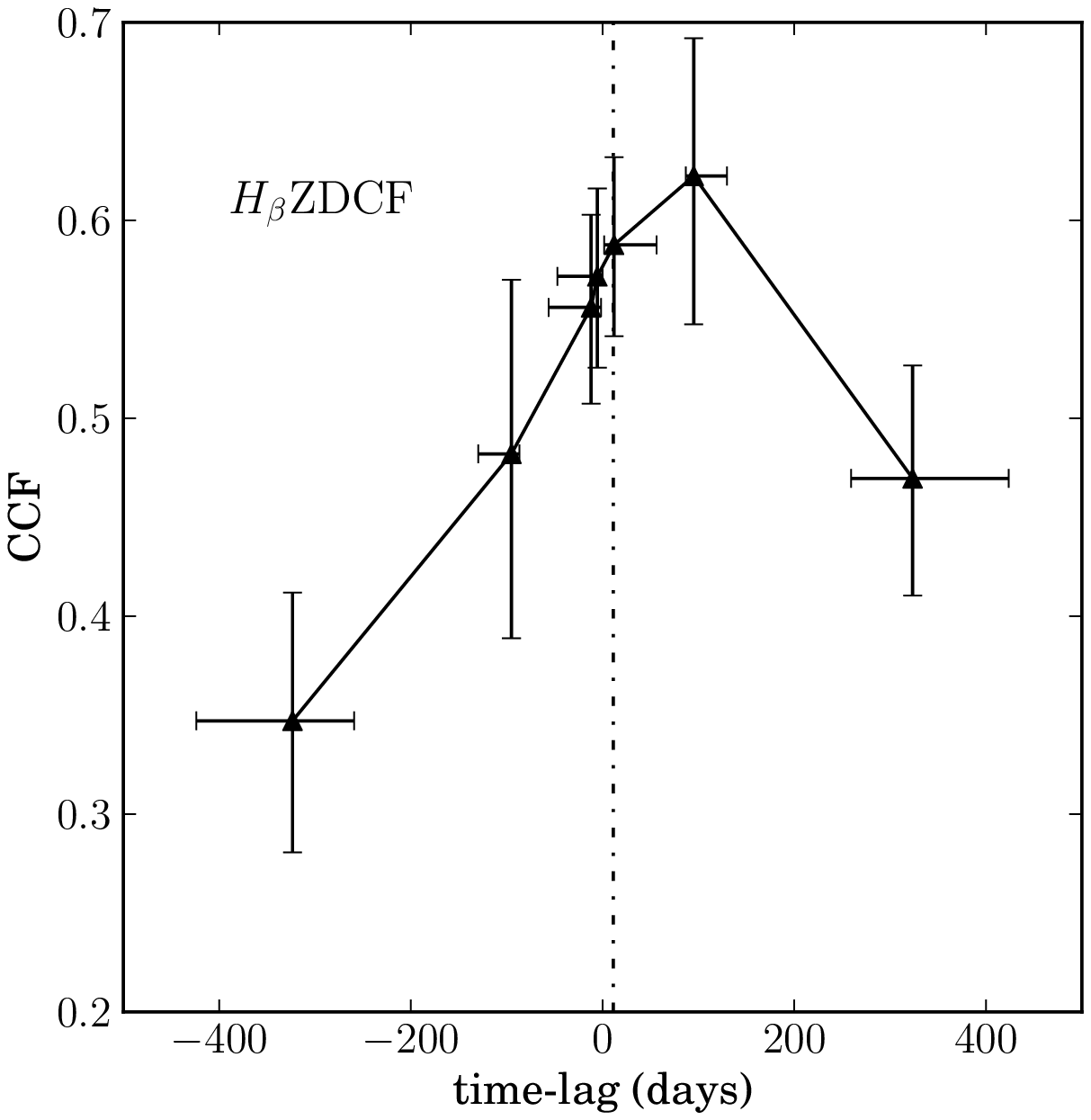}
\includegraphics[width=5cm]{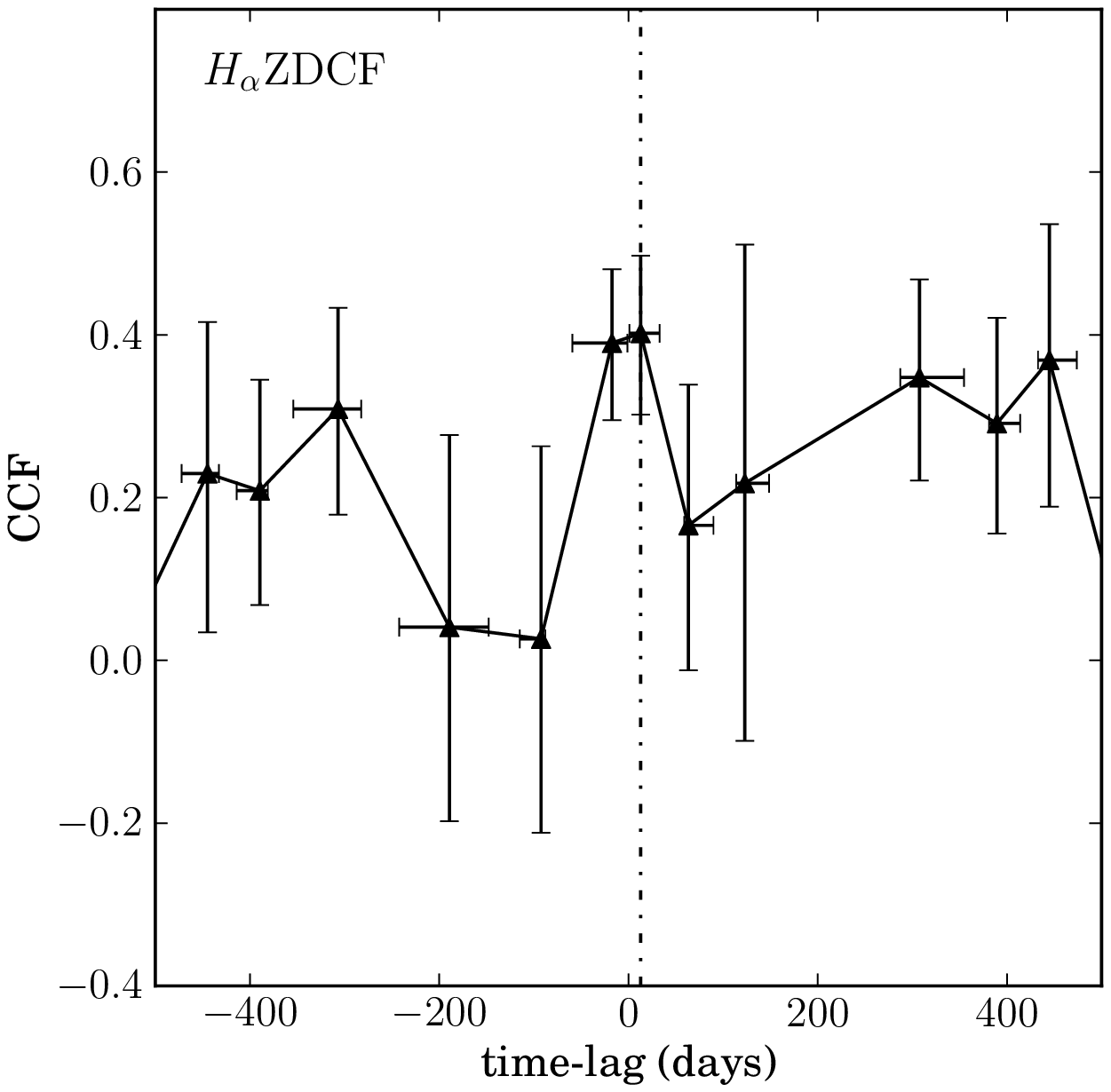}
\includegraphics[width=5cm]{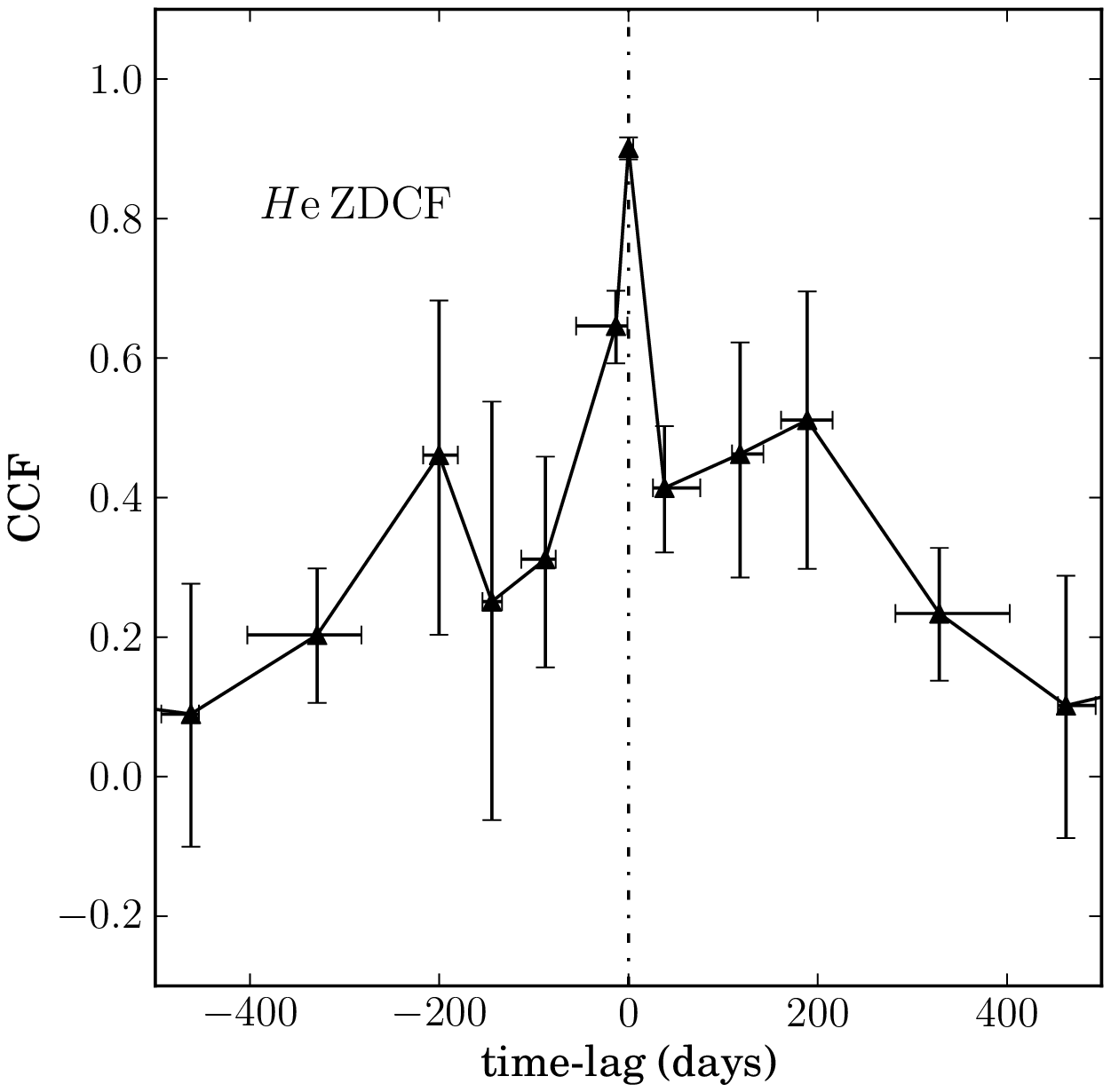}
\includegraphics[width=5cm]{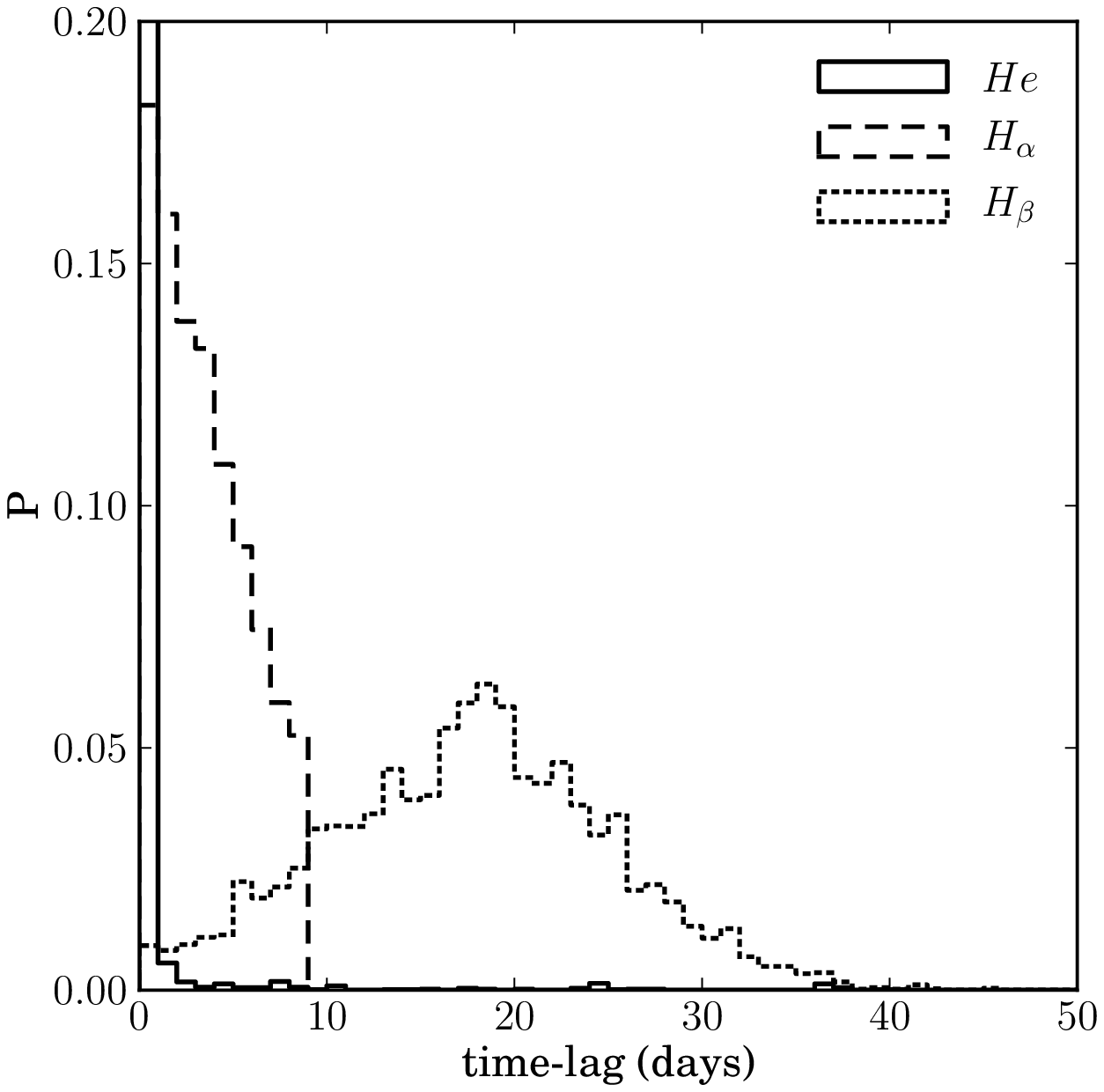}
\caption{Cross correlation functions for the  H$\beta$, H$\alpha$ and He II broad lines.
The upper panels show the ZDCF for H$\alpha$, H$\beta$ and HeII, respectively, while on the bottom 
panel, the SPEAR for three lines is plotted.} \label{ccf}
\end{figure*}

The obtained time-delays are presented in  Fig.~\ref{ccf}. First we find time-delays using the observed data, and we obtained that ZDCF gives the similar lags for H$\alpha$ and H$\beta$ 
around 11--12 days, while SPEAR gives larger differences between H$\alpha$ ($\sim$ 4 days) and H$\beta$ ($\sim$ 17 days).  An important result of the SPEAR analysis is that the probabilities of the
SPEAR time-delays of both curves (see Fig.~\ref{ccf}, right panel) clearly show  unique peaks. 
This fact supports the time-lag range of  11--17 days for 
H$\beta$, and  about 4 times shorter time-lag of H$\alpha$.
From the other hand, the ZDCF and SPEAR time-lags of the He II $\lambda$4686\AA\ line are in a good
agreement with each other as well as with \citet{pet14}, giving time-delays of 2--3 days.

\begin{figure}
\centering
\includegraphics[width=\columnwidth]{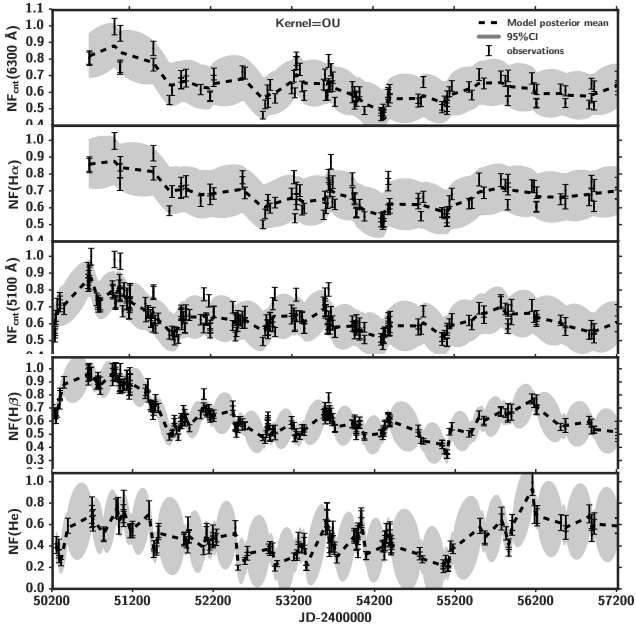}
\caption{Generated light-curves (dashed line) for the continua and broad lines using the Gaussian process regression. Shaded bands represent the 95\% confidence interval (CI) for the predicted light-curves.} \label{lc-gp}
\end{figure}

Since we have long-term observations covering 20 years and observational gaps,
we performed Gaussian process regression (GPR) to generate the light-curves of lines as it is shown in Fig.~\ref{lc-gp}. We found the  time-lags using both observed and generated light-curves.

From generated light-curves we obtained time delays as following:
for  H$\beta$ time-lag is $23.2_{-15.0}^{+5.2}$ days (SPEAR) and 
$21.0_{-0.0}^{+7.0}$ days { with cross correlation coefficient of 
$0.80^{+0.01}_{-0.01}$} (ZDCF), for H$\alpha$ time-lag is $3.2_{-1.0}^{+0.7}$ days (SPEAR) and 
$3.3_{-3.3}^{+9.8}$ days { with cross correlation coefficient of 
$0.80^{+0.01}_{-0.01}$} (ZDCF), and for He II time-lag is $2.5_{-1.5}^{+3.2}$ days (SPEAR) and 
$3.5_{-3.5}^{+10.5}$ days { with cross correlation coefficient of 
$0.44^{+0.03}_{-0.03}$} (ZDCF).

{ To better assess how the time-lag between the continuum and H${\beta}$ line
has behaved during the monitored period, we have divided these light curves into
two halves (according to the length of the monitored period, the dividing date is JD
2453734.63), and determined time-lags for each of them. 
For the first part we obtain the ZDCF time-lag of $15^{+6}_{-1}$ days with cross
correlation coefficient  of  $0.80^{+0.02}_{-0.02}$, and the SPEAR time-lag is
$19.4^{+5.0}_{-4.5}$ days. For the second part the ZDCF time-lag is $28^{+7}_{-1}$ with cross correlation
coefficient of $0.60^{+0.04}_{-0.03}$ and the SPEAR time-lag is $22.0^{+14.2}_{-9.1}$.
This illustrates that the time-lag for H$\beta$  remains consistent 
(within the frame of the error-bars) across the monitored period.}

Our time-delays  for the H$\beta$ and He II lines are quite similar to the ones
obtained by \citet{pet14} and, moreover  they are within  $3\sigma$ 
distance (they found time delay of 11--20 days for H$\beta$ and 1--2 days for He II).
However, it is interesting that the H$\alpha$ line has a smaller time-lag 
than the H$\beta$, and that it is closer to the time-lag of the He II line.
{ This is quite unusual and it seems also unphysical due to following reasons: (i) the
H$\alpha$ and H$\beta$ should come from the same emitting gas, 
and there can be a slight difference between the H$\alpha$ and H$\beta$ lags,
but as a rule the  H$\alpha$ lag is larger than H$\beta$ one \citep[see][]{sh08,sh10}, however the
difference should not be too large as in this case;  (ii) as it is shown in Section 3.2.1 the H$\alpha$
and H$\beta$ have practically the same FWHM and their different lags indicate that the emitting
gas does not follow motion as $r^{-1/2}$, i.e. the gas virialization is not present, that is unlikely to happen.
Additionally, note here again, that H$\alpha$ is more sensitive to the slit offset than H$\beta$. }
{ Finally, the correlation peaks for both H$\alpha$ and He II are not convincing for both used methods
(see Fig.~\ref{ccf}) that have the influence on the lag uncertainties for these lines. Therefore,
considering the above discussion the obtained time-lags for H$\alpha$ and He II (since we have Fe II contribution) 
should be taken with caution.}


\begin{figure}
\centering
\includegraphics[width=\columnwidth]{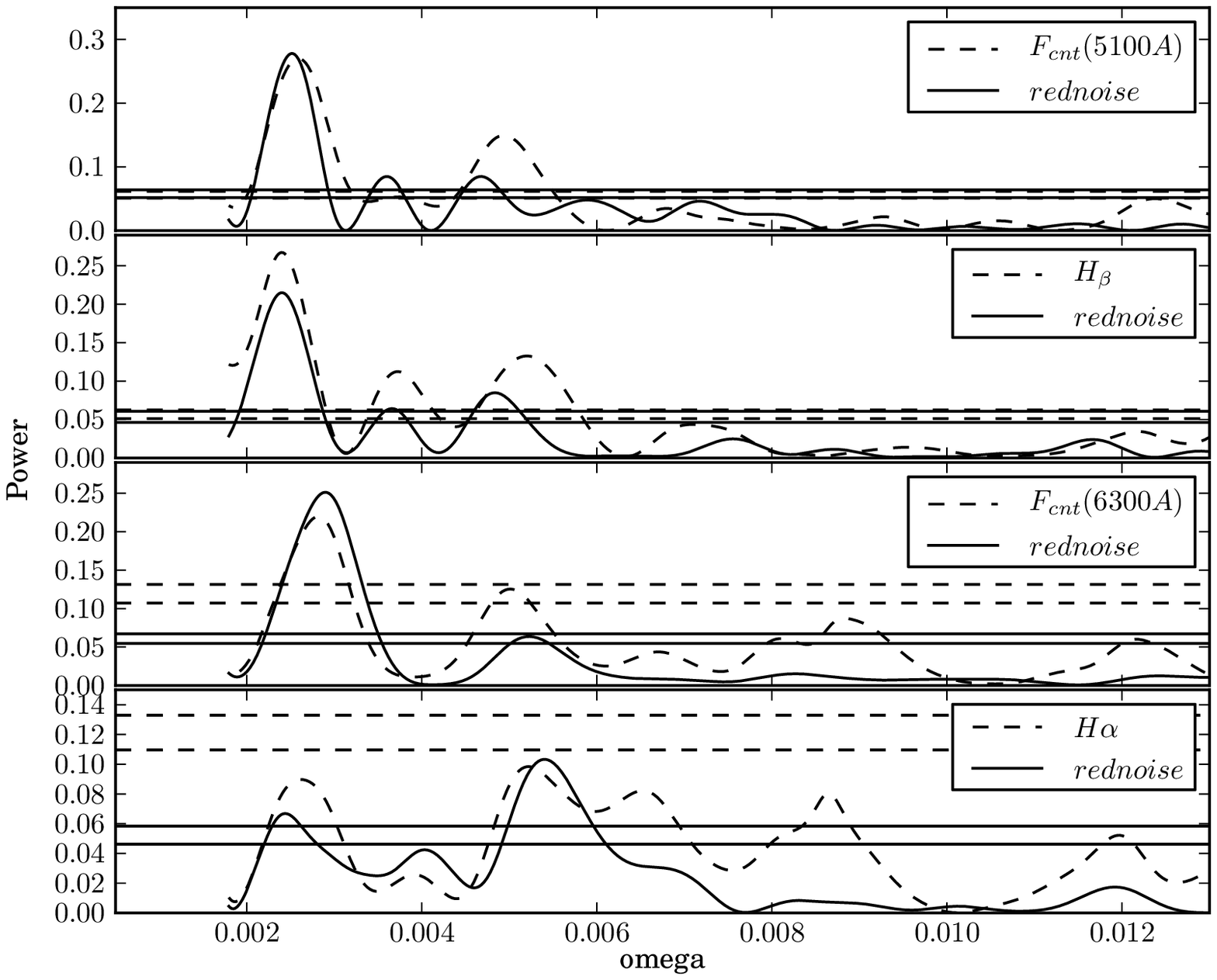}
\caption{Lomb--Scargle periodograms for continua and broad lines original light-curves (dashed lines) 
and simulated red-noise light-curves (solid lines).} \label{period}
\end{figure}

\subsubsection{Periodicity}
\label{sec:period}

A very important and common variability feature in AGN light-curves is red
noise, which can lead to relatively dominating peaks in the power spectrum of  periodograms at
low frequencies. Therefore, it is important to consider possible sources of noise
when testing for periodicity in light-curves. Although this type of  flux variability is
called "red noise", this variability most likely  arises from physical processes from the
source.

Therefore, in our timing analysis we tested whether a purely red
noise model can produce periodic variability of light-curves. For this
purpose, all red noise simulations were produced by Ornstein--Uhlenbeck
(OU) process (red noise) within the framework of
Gaussian process regression model \citep[][]{rw06}.
We simulated OU process with a variance of real light-curves.
Then we obtained random light-curves from the OU process sampled to a
regular time interval. We used them to test that  a periodicity of the observed light
curves is the product of random variations.

For each observed and modeled light-curve we calculated Lomb--Scargle
periodogram with a bootstrap analysis  to compute its significance as it
is described in \citet{sh16}. 

The periodogram analysis (Fig \ref{period})) shows that
there are three possible periods of $\sim$2600, 1700, and 1200 days
present in observed data. However, in the H$\alpha$ line, these peaks are not
significant, while in the case of the continuum at 6300 \AA, the peak of 1700 days
is missing. Perhaps this period is not relevant. Moreover, these peaks are
presented in random walk simulated data and their power is comparable to those in the observed data,
especially in the case of the continuum at 5100 \AA\, (Fig \ref{period}).  
Only in the case of H$\beta$, two smaller peaks are of grater power then those in red noise
 simulated data. 
 
Therefore, the possible periods are 2600 and 1200 days, but they must be
interpreted with caution because similar periodicities are also prominent in the red noise simulations.
Thus, we propose that the detected periodicities are the product of either:
i) random variations within the object; ii)  the sampling resolution; iii)
some periodic process within the core of  object, or iv) a combination of all above factors.

\section{Discussion}

We explore the long-term photometric and spectroscopic variability of the NGC 7469 AGN observed with four telescopes 
in a common monitoring campaign that lasted for 20 years (from 1996 to 2015). We carefully selected the observed spectra, 
taking into account that there may be a slit offset during the observations, in which case the contribution of the stellar circumnuclear ring can affect the 
continuum and line fluxes. Our measurements are presented in Tables \ref{tab1} and \ref{tab5}.

The obtained results can be compared and discussed in the frame of previous findings. As we noted above,
NGC 7469 is one of the most observed and investigated type 1 AGNs, from early 60's and 70's 
\citep[see e.g.][etc.]{bu63,ul72,ya73,an73,pr74,pr75,pr76}. First spectral monitoring campaigns started at the middle of 70's \citep[see][]{pr74,pr75,pr76} with the aim to investigate changes in the broad lines (mostly in the H$\beta$ spectral range) and 
continuum. First spectral observations of NGC 7469 showed rapid variations, with the broad line component almost disappeared
\citep{ch90,pr96,pro97}. However, these results have to be take with caution, since we demonstrated that the slit offset can significantly affect the observed broad line intensities (see Section~\ref{sec:flux}). 

On the other side, the rapid photometric variability of the nucleus seems to be present and observed 
\citep[see][]{dul92,ly95,gl98,mer00,art10}, very often as the short-term outburst, which are in support of our finding of flare-like events seen in the photometric and spectroscopic
ligh-curves (see Section~\ref{sec:flare}).

Reverberation monitoring campaigns (in the X, UV and optical band) started at the end of 90's \citep{wan97,col98,nan98}, 
in which the time delay of 
different broad lines have been determined, and some estimates of the central black hole mass were given.
The radio-band monitoring for eight years was performed by \cite{per09}, but was more concentrated on 
supernova events in the circumnuclear ring. It is interesting that we could not find significant changes in the
optical photometric and spectroscopic light-curves after the explosion of SN 2000ft \citep{col01,col07} (see Fig.
\ref{fig2}). In this period we could not detect an outburst or flare like events.

\begin{figure}
\centering
\includegraphics[width=8cm]{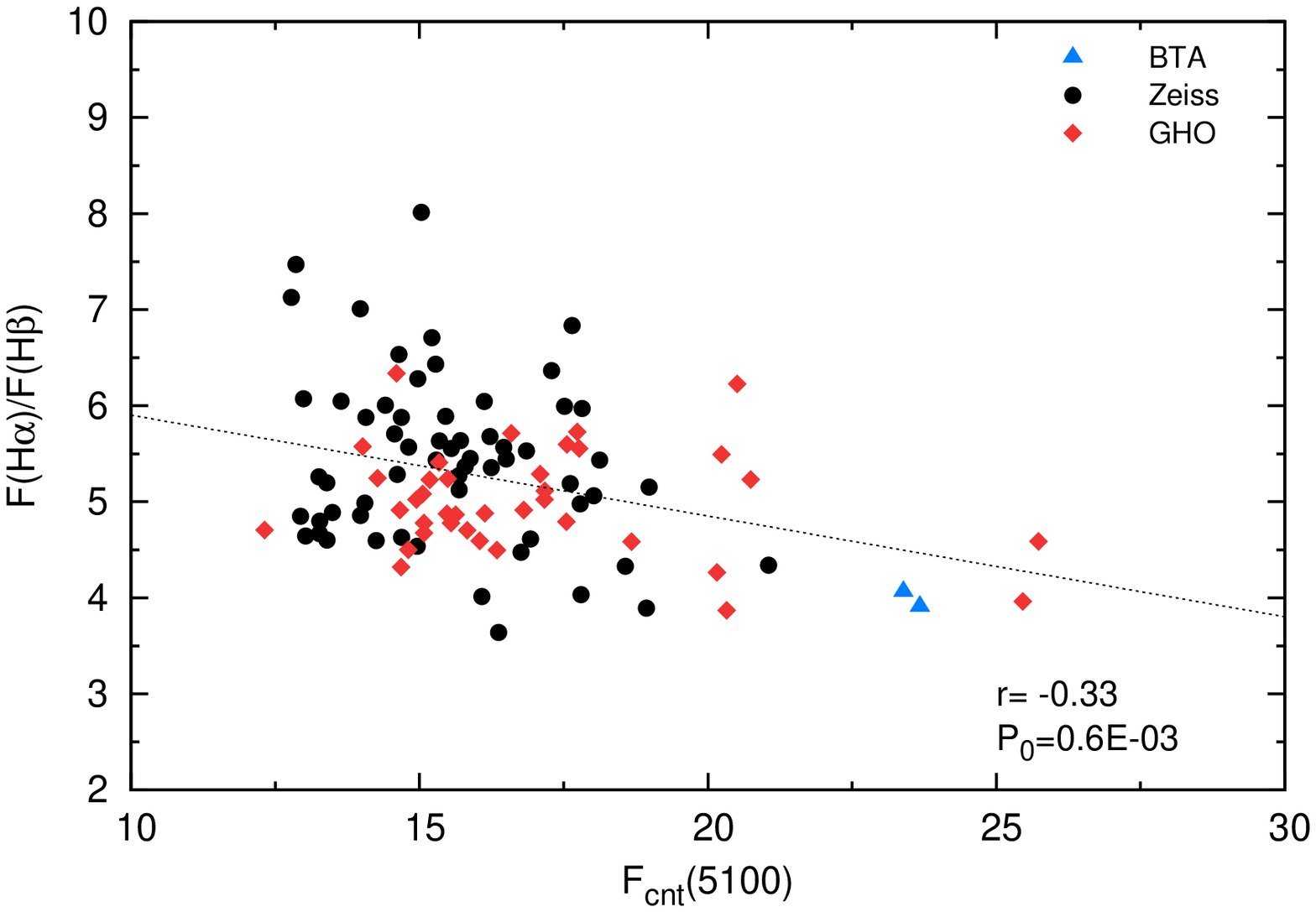}
\includegraphics[width=8cm]{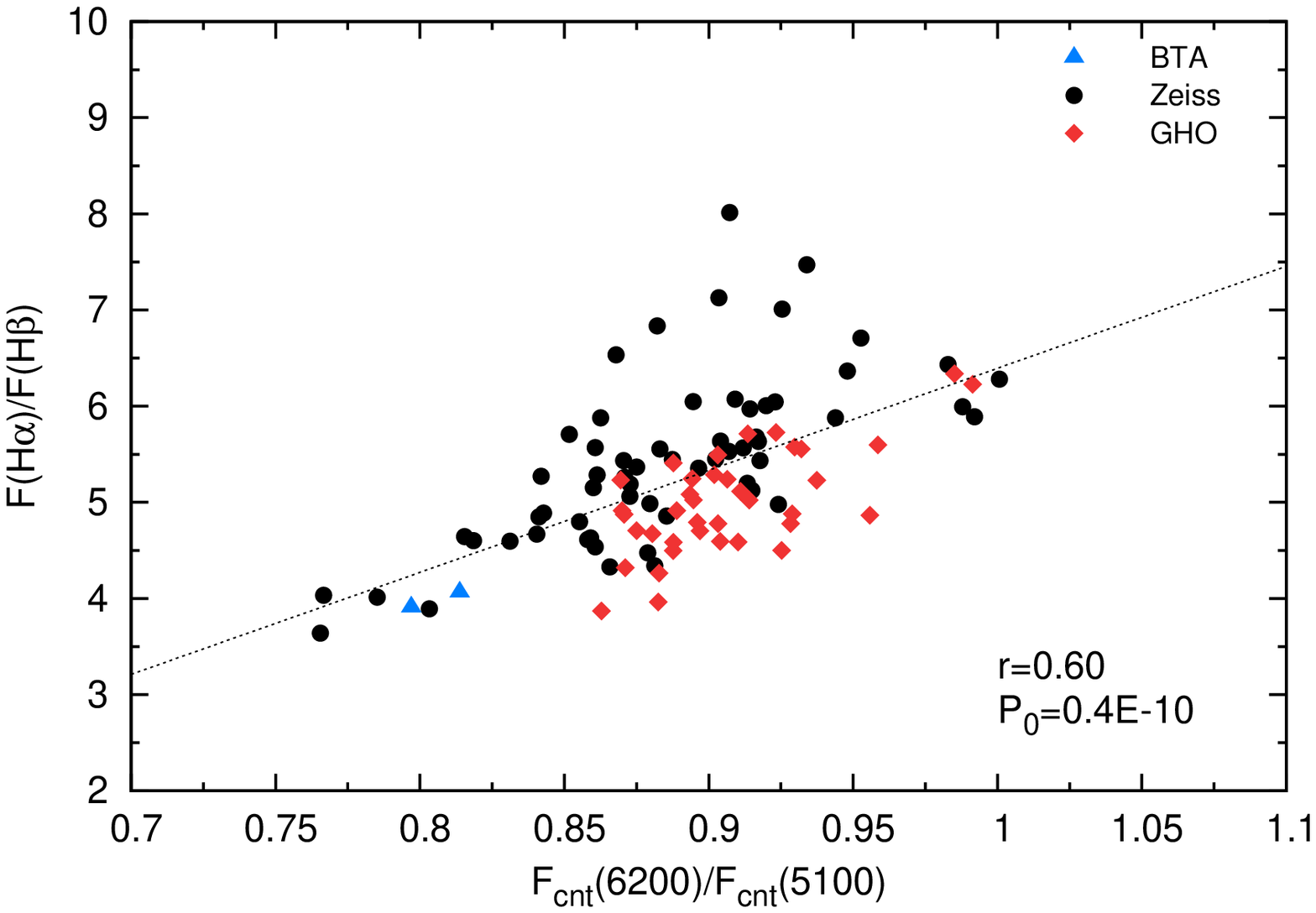}
\caption{The H$\alpha$/H$\beta$ flux ratio vs. continuum at 5100 \AA\ (upper) and the H$\alpha$/H$\beta$ flux 
ratio vs. 6200\AA/5100\AA\ continuum flux ratio (bottom). Symbols are the same as in Fig.~\ref{fig2}.} \label{cnt-ratio}
\end{figure}

\begin{figure}
\centering
\includegraphics[width=9cm]{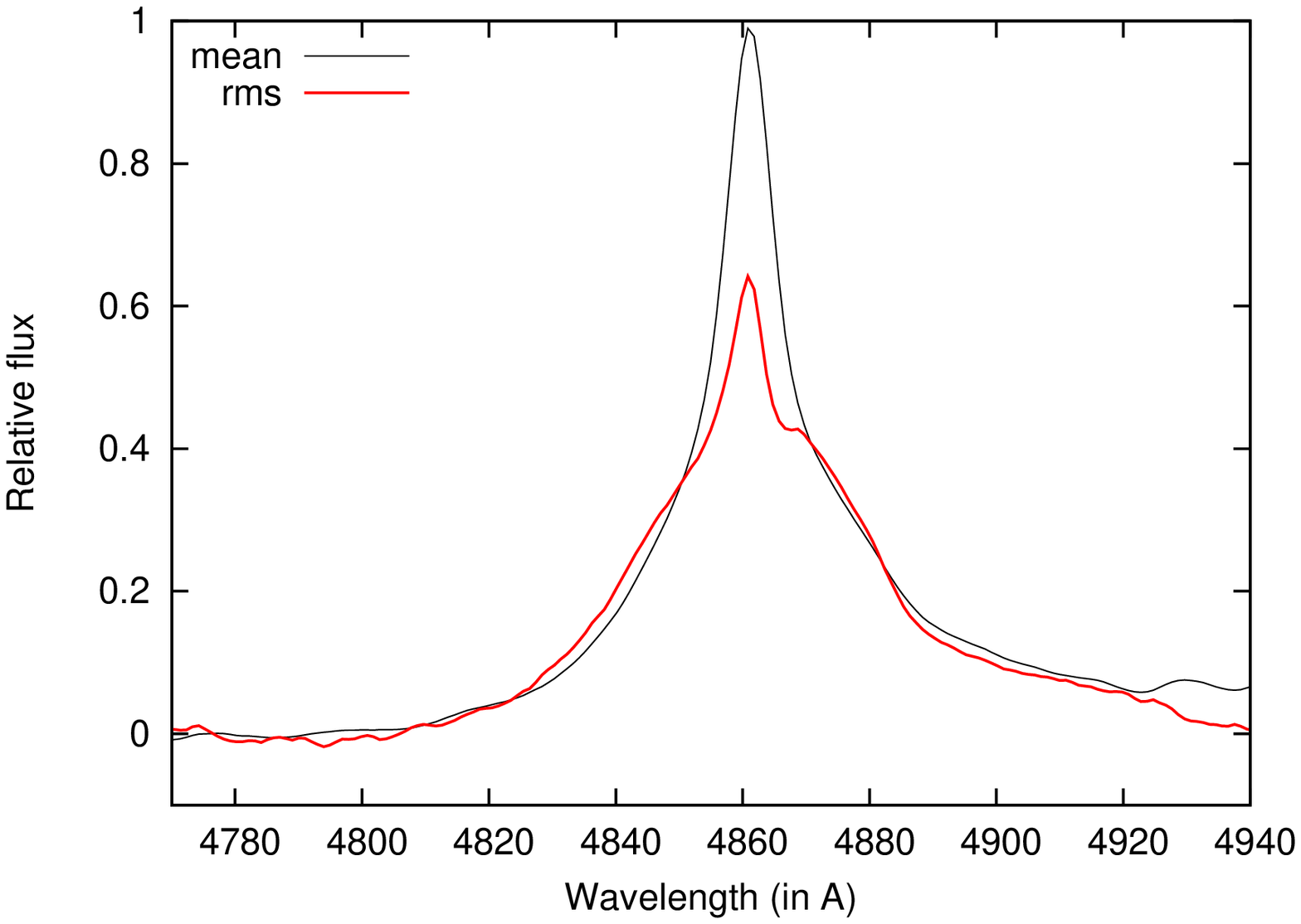}
\caption{Comparison between the mean H$\beta$ line and its rms, both profiles are normalized to the line wings.
} \label{mean-rms}
\end{figure}

\subsection{Long-term variability -- nature of the BLR and continuum source}

We performed different CCF analyses to find the time delays between the continua and broad emission lines. The analysis of the obtained results and measurements allows us to discuss the nature of the BLR of NGC 7469.

We found that there are very strong correlations between the H$\beta$ and H$\alpha$ line flux variabilities with the corresponding continuum (at 5100 \AA\ and 6300 \AA\, respectively).  This is in favour  of the classical view of the BLR which is composed from the photoionized gas. Additionally, it was shown before that the UV emission is caused by the reprocessing of the X-rays \citep[see][]{pet04}. Also, we show in Fig.~\ref{cnt-ratio} the H$\alpha$/H$\beta$ ratio as a function of the continuum at 5100 \AA\ and the 6200\AA/5100\AA\ continuum flux ratio. Fig.~\ref{cnt-ratio} shows that the H$\alpha$/H$\beta$ ratio is decreasing with the continuum flux at 5100 \AA, but it is increasing with the  6200\AA/5100\AA\ continuum flux ratio. This indicates that the broad line ratio are sensitive to the continuum flux, and therefore, the gas in the BLR is  most likely photoionized by the central continuum source (accretion disc).

On the other hand, there is a smaller correlation between the flux variation in the He II $\lambda$4686\AA\ line and the continuum at 5100 \AA\ than in Balmer lines. In principle it can be caused by uncertainties in
the He II line flux measurements, since the line is very weak (especially in the low-activity AGN state) and blended with Fe II lines.

Although there is a good correlation between the H$\alpha$ and H$\beta$ flux variability, the CCF 
analysis showed significantly different time delays, which are in principle connected with the BLR dimensions.
It seems that the H$\beta$ emission region is significantly larger than the H$\alpha$ one. 
This should be taken with caution, since the H$\alpha$ broad line is very sensitive to  the slit offset, 
(see Fig.~\ref{slit} and discussion in Section~\ref{sec:flux}). I.e.
a small slit offset of 1\arcsec\ has high influence to the normalized broad H$\alpha$ flux, but does not significantly affect the normalized H$\beta$ flux. Therefore, the lag obtained for H$\beta$ seems to be more relevant for discussion of the BLR dimension of around 20 light days.
The He II has  small time-lag (2--3 days) that is  smaller than H$\beta$, which is usually detected in Type 1 AGNs. 
{ The obtained time-lags for He II and H$\beta$ are in a good agreement with previous 
measurements, as e.g. \citet{pet14} found lags of 11.8$^{+4.2}_{-2.6}$
for H$\beta$, and 1.9$^{+7.5}_{-1.6}$ for He II using the CCF, and
23.6$^{+1.8}_{-2.7}$ for H$\beta$, and 10.4$^{+2.1}_{-0.7}$ for He II 
using the SPEAR method (see Table 4 in their paper). I.e. both
methods gave that the He II line emitting region is smaller than the 
H$\beta$ one, this is in agreement with the line widths, since
the He II FWHM seems to be larger than the H$\beta$ one.}

From this long-term monitoring of NGC 7469, we can conclude that the geometry of the BLR was not changed.
As an illustration, in Fig.~\ref{mean-rms} we compared the mean and rms H$\beta$ line profiles, rescaling the rms wing intensities to the ones of the mean profile. Fig.~\ref{mean-rms} shows that the broad components are the same (there are some residuals in the centre of the rms caused by small differences in the spectral resolution). There is a slight red asymmetry, but there are no significant changes in the broad 
line profiles during the 20-year long monitoring period. From Gaussian fitting, shown in Fig.~\ref{broad}, we found that the broader Gaussian (fitting the far red wing) is shifted for about 600 km s$^{-1}$ to the red. This asymmetry might be due to some kind of inflow, but taking that there is no changes in this (red) part of the broad line profiles during 20-year period, the inflow should be very stable or with large dimension.

\subsection{Black hole mass}

The results of the long-term monitoring can be used for the black hole mass estimation ($M_{BH}$) as
\begin{equation}
    M_{\rm GRAV}=f\cdot M_{\rm vir},
	\label{eq:mass}
\end{equation}
$f$ is a factor that depends on the geometry of the BLR and $M_{\rm vir}$ is the so called "virial product" or Keplerian mass \citep[see][]{col98,pet14} that can be calculated as:
\begin{equation}
    M_{\rm vir}={V_{\rm FWHM}^2\cdot R_{\rm BLR}\over G},
	\label{eq:mass_vir}
\end{equation}
where $V_{\rm FWHM}$ is the  orbital velocity at that radius ($R_{\rm BLR}$) of the BLR. 

We found that the lag for H$\beta$ is around 20 days and for H$\alpha$ around 3 days (see Section~\ref{sec:lag}), the corresponding FWHMs of these two lines are 2000 km s$^{-1}$ and 2100  km s$^{-1}$, respectively. The FWHMs are measured from the broad components only (see Section~\ref{sec:var}), then for the approximation of a spherical geometry in the BLR, the orbital velocity vs. FWHM relationship can be written as \citep[see][]{col98} :
\begin{equation}
    V=(\mathstrut{\sqrt{3}}/2)\cdot {\rm FWHM}.
	\label{eq:vel}
\end{equation}
The orbital velocity from the H$\beta$ FWHM is $\sim$1700 km s$^{-1}$, and slightly higher for  H$\alpha$, V$\approx$1800 km s$^{-1}$.

We calculated the virial product using the following relation (after including constants and unit transformations):
\begin{equation}
    M_{\rm vir}[M_\odot] \approx 0.197\cdot V[{\rm km\ s^{-1}}]\cdot R_\tau[{\rm l.d.}],
	\label{eq:mass_vir2}
\end{equation}
where  $R_\tau=c\tau$ is determined from the time delays in H$\alpha$ and H$\beta$.

The obtained virial black hole mass indicators are: $2\cdot 10^6M_\odot$ obtained from the H$\alpha$ FWHM and time-lag and $1.1\cdot 10^7M_\odot$ obtained from the H$\beta$ line parameters. The obtained result for H$\alpha$ is in agreement with the one given in \citet{pet14}, however, the obtained mass from
H$\beta$ seems to be bigger than given in \cite{col98,pet14}. This is obviously
because of our estimates of the H$\beta$ BLR dimension, for which we found that is around 20 ld. However, 
the difference is not so big, and taking that the factor $f$ might have value from 1. to 5.5 \citep{on04}, the estimated black hole mass is around $\sim (1-6)\cdot 10^7M_\odot$ that is in the agreement with previously given estimates \citep[][]{col98,pet14}.



\section{Conclusions}

We have presented the long-term photometric and spectroscopic monitoring campaign for NGC 7469. Previous spectral monitoring campaigns of this galaxy were shorter, and it is for the first time that the data of the 20-year long monitoring of this AGN are presented. Additionally, we present the results
of the 16--year period of photometric monitoring. We measured the continuum and line fluxes (continua at 5100 \AA\ and 
6200 \AA, and H$\alpha$, H$\beta$ and He II $\lambda$4686\AA\ lines). The measured data have been analysed and different correlation between the line and continuum fluxes have been explored. We determined the time delays between the continuum and line flux variations,
and at the end we used measured data for discussing the nature of the variability, the BLR structure, and to estimate the central black hole mass in NGC 7469.

Concerning our investigations of the long-term variability  of NGC 7469, we can outline the following conclusions:

\begin{itemize}

\item[(i)] The AGN of NGC 7469 is surrounded by complex circumnuclear region, 
that may affect the observed spectra in a monitoring campaign. Our simulations show that the narrow slit observations, with relatively small offset changing (1-2\arcsec) can produce an artificial variability in the line flux, especially in the case of the H$\alpha$ line.
Therefore, for future monitoring campaigns one should have this in mind.
 
 \item[(ii)] The long-term variability shows that the broad line and continuum  fluxes of the 
 AGN have changed for around  $F_{max}/F_{min}\sim2$, in the continuum and H$\alpha$ line, while in 
 the H$\beta$ it changed for around 3 times.  The changes in the He II line seem to be around 6 times.
 We found that in the flux variability two periods around 2600 and 1200 days may be present, however these
 periods have to be taken with caution since  red noise may be also present. We confirm that there are several 
 short time (1--5 days) flare-like peaks in the spectral light-curves.
 
 \item[(iii)] The lags of the H$\beta$ and H$\alpha$ broad lines are quite different,
 showing that the H$\alpha$ broad line emitting region (similar as He II) is around $\sim2-3$
 light days, but H$\beta$ broad lines is coming from a region that is ten times larger ($\sim$ 20 
 light days). As we noted above, the lag for H$\alpha$ should be taken with caution.
 
 \item[(iv)] The broad line profiles did not show large change, and it seems that the emission regions are geometrically stable. The H$\beta$ and H$\alpha$ have FWHM around 2000 km s$^{-1}$ and show a red asymmetry. This red asymmetry indicates some kind of the receding motion of the BLR emission gas with velocity of $\sim 600$ km s$^{-1}$. It is  interesting that there is no changes in the red asymmetry in the 20-year period.
 \end{itemize}
 
We estimated the mass of the central black hole, taking the rotation velocity from the broad H$\beta$ and H$\alpha$ line FWHMs and their time delay, and found that the mass is  $\sim(1-6)\cdot 10^7M_\odot$, which is in an agreement with previous estimates from shorter { but more intensive} monitoring campaigns.

\section*{Acknowledgements}

This work was supported by: INTAS (grant N96-0328), RFBR (grants
N97-02-17625 N00-02-16272, N03-02-17123, 06-02-16843, N09-02-01136,
12-02-00857a, 12-02-01237a,N15-02-02101), CONACYT research grants 39560, 54480,
and 151494, and the Ministry of Education and Science of Republic of Serbia through the project
Astrophysical Spectroscopy of Extragalactic Objects (176001). 
L. \v C. P., W. K. and D. I. are grateful to the Alexander von Humboldt
foundation for support in the frame of program "Research Group
Linkage". W. K. is supported by the DFG Project Ko 857/32. 
We especially thank Borisov N.V., Fathulin T., Fioktistova I., Moiseev A., 
Mikhailov V. and Vlasyuk V.V. for taking part in the observations.












\bsp	
\label{lastpage}
\end{document}